\documentclass[aps,prd,print,groupedaddress,nofootinbib,eqsecnum,onecolumn]{revtex4}
\usepackage{eurosym}
\usepackage{amsmath}
\usepackage{amsthm}
\usepackage{amssymb}
\usepackage{graphicx}
\usepackage{slashed}
\usepackage{hyperref}
\usepackage{epstopdf}

\theoremstyle{plain}

\theoremstyle{definition}

\begin{document}

\author{Alessio Marrani}
\email{jazzphyzz@gmail.com}
\affiliation{Instituto de F\'{\i}sica Teorica, Dep.to de F\'{\i}sica,\\
Universidad de Murcia, Campus de Espinardo, E-30100, Spain}
\author{Michael Rios}
\email{mrios@dyonicatech.com}
\affiliation{Dyonica ICMQG, Los Angeles, CA, USA}
\author{David Chester}
\email{davidC@quantumgravityresearch.org}
\affiliation{Quantum Gravity Research, Los Angeles, CA, USA}
\title{Monstrous M-theory}

\begin{abstract}
In $26+1$ space-time dimensions, we introduce a gravity theory whose
massless spectrum can be acted upon by the Monster group when reduced to $%
25+1$ dimensions. This theory generalizes M-theory in many respects and we
name it Monstrous M-theory, or M$^{2}$-theory. Upon Kaluza-Klein reduction
to $25+1$ dimensions, the M$^{2}$-theory spectrum irreducibly splits as $%
\mathbf{1}\oplus\mathbf{196,883}$, where $\mathbf{1}$ is identified with the
dilaton, and $\mathbf{196,883}$ is the dimension of the smallest non-trivial
representation of the Monster. This provides a field theory explanation of
the lowest instance of the Monstrous Moonshine, and it clarifies the
definition of the Monster as the automorphism group of the Griess algebra,
by showing that such an algebra is not merely a sum of unrelated spaces, but
descends from massless states for M$^{2}$-theory, which includes Horowitz
and Susskind's bosonic M-theory as a subsector. Further evidence is provided
by the decomposition of the coefficients of the partition function of
Witten's extremal Monster SCFT in terms of representations of $SO_{24}$, the
massless little group in $25+1$; the purely bosonic nature of the involved $%
SO_{24}$-representations may be traced back to the unique feature of $24$
dimensions, which allow for a non-trivial generalization of the triality
holding in $8$ dimensions. Last but not least, a certain subsector of M$^{2}$%
-theory, when coupled to a Rarita-Schwinger massless field in $26+1 $,
exhibits the same number of bosonic and fermionic degrees of freedom; we
cannot help but conjecture the existence of a would-be $\mathcal{N}=1$
supergravity theory in $26+1$ space-time dimensions. \newline
\end{abstract}

\maketitle

\bigskip

\bigskip

\begin{center}
\textit{Dedicated to John H. Conway}
\end{center}

\newpage

\tableofcontents

\newpage

\setcounter{page}{1}

\section{Introduction}

The Monster group $\mathbb{M}$, the largest of sporadic groups, was
predicted to exist by Fischer and Griess back in the mid 70's \cite{griess76}%
. $\mathbb{M}$ is the automorphism group of the Griess algebra, as well as
the automorphism group of the Monster vertex operator algebra (VOA) \cite%
{flm,conway85}. Conway and Norton defined \textit{Monstrous Moonshine} as
the observation that the Fourier coefficients of the $j$-function decompose
into sums of dimensions of representations of $\mathbb{M}$ itself \cite%
{conway79} and this was proven by Borcherds using generalized Kac-Moody
algebras \cite{borcherds}. In the language of conformal field theory (CFT),
Monstrous Moonshine is the statement that the states of an orbifold theory,
which is the $D=25+1$ bosonic string theory on $(\mathbb{R}^{24}/\Lambda
_{24})/\mathbb{Z}_{2}$ (where $\Lambda _{24}$ is the Leech lattice \cite%
{conwaybook,wilsonLeech,uleech}), are organized in representations of the
Monster group, with partition function equivalent to the $j$-function \cite%
{harveyb,bimonster,mbps}. Witten also found the Monster group in
three-dimensional pure gravity \cite{witten3d}, for $AdS_{3}$, where the
dual CFT is expected to be that of Frenkel, Lepowsky and Meurman (FLM) \cite%
{flm}. A Monster SCFT and fermionization of the Monster CFT were also
defined and studied \cite{harveyb,fmonster}

Eguchi, Ooguri and Tachikawa later noticed the elliptic genus of the K3
surface has a natural decomposition in terms of dimensions of irreducible
representations of the largest Mathieu group $M_{24}$ \cite{eguchi}, and
this was named \textit{Umbral Moonshine} \cite{umbral1,umbral2}, which
generalizes the Moonshine correspondence for other sporadic groups \cite%
{sporadicft}.

With Witten's proposal \cite{witten95} that M-theory unifies all the
ten-dimensional string theories with $\mathcal{N}=1$ supergravity in $D=10+1$
space-time dimensions, Horowitz and Susskind argued \cite{bMtheory} there
exists a bosonic M-theory in $D=26+1$ that reduces to the bosonic string in $%
25+1$ upon compactification. As the Monster group has a string theoretic
interpretation in $D=25+1$ \cite{bimonster,moonmod,genusz}, it is also
natural to consider its action on fields from $D=26+1$; support for this is
found from bosonic M-theory's M2-brane near horizon geometry $AdS_{4}\times
S^{23}$, discussed by Horowitz and Susskind as an evidence for a dual $2+1$
CFT with global $SO_{24}$ symmetry \cite{bMtheory}. By observing that the
automorphism group of the Leech lattice $\Lambda _{24}$, the Conway group $%
Co_{0}$ \cite{conwaybook,wilsonLeech}, is a maximal finite subgroup of $%
SO_{24}$, and its $\mathbb{Z}_{2}$ quotient $Co_{1}\simeq Co_{0}/\mathbb{Z}%
_{2}$ is a maximal subgroup of the Monster \cite{conwaybook,flm}, it is
possible to realize some Monstrous symmetry as a finite subgroup of $%
\mathcal{R}$-symmetry in $26+1$ dimensions \cite{wmtheory}.\medskip

In the present paper, we introduce an Einstein gravity theory coupled to $p$%
-forms in $26+1$ space-time dimensions, which contains the aforementioned
bosonic string theory \cite{bMtheory} as a subsector. We name such a theory
\textit{Monstrous M-theory}, or shortly \textit{M}$^{2}$\textit{-theory},
because its massless spectrum (with gauge fields mod $\mathbb{Z}_2$) has the
same dimension ($196,884$) as the Griess algebra and upon dimensional
reduction can be acted upon by the Monster group $\mathbb{M}$ itself. When
reducing to $25+1$, a web of gravito-dilatonic theories, named \textit{%
Monstrous gravities}, is generated, in which the decomposition $196,884=%
\mathbf{196,883}\oplus \mathbf{1}$, which first hinted at Monstrous
Moonshine \cite{conway79}, entails the fact that the dilaton scalar field $%
\phi $ in $25+1$ is a singlet of $\mathbb{M}$ itself. As such, the
irreducibility under $\mathbb{M} $ is crucially related to dilatonic gravity
in $25+1$ space-time dimensions. The existence of a \textquotedblleft
weak\textquotedblright\ form of the $SO_{8}$-triality for $SO_{24}$, which
we will name $\mathbf{\lambda }$-triality, gives rise to a $p$($\geqslant 0$%
)-parametrized tower of \textquotedblleft weak\textquotedblright\ trialities
involving $p$-form spinors in $24$ dimensions, which we will regard as
massless $p$-form spinor fields in $25+1$ space-time dimensions. Such
\textquotedblleft weak\textquotedblright\ trialities are instrumental to
provide most of the Monstrous gravity theories with a fermionic (massless)
spectrum, as well, such that the spectrum is still acted upon by the Monster
$\mathbb{M}$.

All this gives an elegant description of the Monster's minimal non-trivial
representation $\mathbf{196,883}$, in relation to the total number of
massless degrees of freedom of Monstrous gravities in $D=25+1$; as such,
this also elucidates the definition of $\mathbb{M}$ as the automorphism
group of the Griess algebra (the degree two piece of the Monster VOA), which
has been considered artificial in that it was thought to involve an algebra
of two or more unrelated spaces \cite{flm,conwaybook,borcherds2002}.\bigskip

The plan of the paper is as follows. We give motivation for Monstrous
M-theory by lifting the M2-brane from $D=10+1$ to $D=26+1$ and breaking the
Poincar\'{e} symmetry in its near-horizon geometry, which results in an $%
SO_{24}$ $\mathcal{R}$-symmetry, that has the Conway group $Co_{0}$ as a
maximal finite subgroup. We then reduce the near-horizon geometry of the
M2-brane in $D=26+1$ and relate the holography to Witten's BTZ black hole
\cite{witten3d} with Monstrous symmetry. Next, in Sec. \ref{triality} we
briefly review the \textit{triality} among the $8$-dimensional
representations of the Lie algebra $\mathfrak{d}_{4}$, and then, in Sec. \ref%
{lambda} we introduce some \textquotedblleft weak\textquotedblright\
generalization for the Lie algebra $\mathfrak{d}_{12}$, which we will name $%
\mathbf{\lambda }$-triality, giving rise to the $\mathbf{\psi }$-triality,
as discussed in Sec. \ref{psi}. As it will be seen in the treatment below,
the \textquotedblleft weakness\textquotedblright\ of the aforementioned
trialities relies on the \textit{reducibility} of the bosonic
representations involved. Then, in Sec.\ \ref{MGra} we introduce and
classify\ non-supersymmetric, gravito-dilatonic theories, named \textit{%
Monstrous gravities}, in $25+1$ space-time dimensions, whose massless
spectrum (also including fermions in most cases) has dimension $196,884$,
namely the same dimension as the Griess algebra \cite{flm,conway85}. A
purely bosonic uplift to $26+1$ space-time dimensions is discussed in Sec. %
\ref{MMtheory}; in this framework, we introduce the \textit{Monstrous
M-theory}, also named \textit{M}$^{2}$\textit{-theory}, and we discuss its
possible Lagrangian in Sec. \ref{Lagr}. Moreover, Sec. \ref{BBFF} discusses
a subsector of the M$^{2}$-theory which diplays the same number of bosonic
and fermionic massless degrees of freedom in $26+1$; in Sec. \ref{SS}, this
allows us to conjecture a Lagrangian and local supersymmetry transformations
for the would-be $\mathcal{N}=1$ Einstein supergravity theory in $26+1$
space-time dimensions. Then, Sec. \ref{Coho} presents a cohomological
construction of both the $\mathfrak{e}_{8}$ root lattice and the Leech
lattice $\Lambda _{24}$ (respectively determining optimal sphere packing in $%
8$ and $24$ dimensions \cite{conwaybook}), and all this is again related to
M-theory (i.e., $\mathcal{N}=1$ supergravity) in $D=10+1$ and to the
aforementioned would-be $\mathcal{N}=1$ supergravity in $26+1$,
respectively. Before concluding the paper, in order to provide further
evidence for a consistent higher-dimensional field theory probed by $\mathbb{%
M}$, we decompose the first coefficients of the partition function of
Monster CFT, firstly put forward by Witten \cite{witten3d}, in terms of
dimensions of representations of $SO_{24}$, namely of the massless little
group in $25+1$ space-time dimensions; an interesting consequence of the
aforementioned \textquotedblleft weak\textquotedblright\ trialities
characterizing $SO_{24}$ is that the relevant $SO_{24}$-representations can
be reduced to be \textit{only} the $p$-form ones, $\wedge ^{p}$, for
suitable values of $p$ and with non-trivial multiplicities. Final comments
are then contained in the conclusive Sec. \ref{Conclusion}. An appendix,
detailing the Chern-Simons Lagrangian terms for M$^{2}$-theory, concludes
the paper.


\section{Evidence for Monstrous M-theory}

\subsection{Bosonic M-theory in $D=26+1$}

Horowitz and Susskind conjectured there exists a strong coupling limit of
bosonic string theory that generalizes the relation between M-theory and
superstring theory, called bosonic M-theory \cite{bMtheory}. The main
evidence for the existence of such a $D=26+1$ theory comes from the dilaton
and its connection to the coupling constant, with the dilaton entering the
action for the massless sector of bosonic string theory as

\begin{equation}
S=\int d^{26}x\sqrt{-g}e^{-2\phi }\left[ R+4\nabla _{\mu }\phi \nabla ^{\mu
}\phi -\frac{1}{12}H_{\mu \nu \rho }H^{\mu \nu \rho }\right] ,
\end{equation}

\noindent in a way similar to type IIA string theory, as if representing the
compactification scale of a Kaluza-Klein reduction from $D=26+1$ space-time
dimensions with $\mathbf{324}\rightarrow\mathbf{299}+\mathbf{24}+\mathbf{1}$
graviton decomposition. However, while in type IIA string theory the
existence of a vector boson in the string spectrum implies an $S^{1}$
compactification, in closed bosonic string theory there is no massless
vector. For this reason, an $S^{1}/\mathbb{Z}_{2}$ orbifold compactification
of bosonic M-theory was proposed as its origin \cite{bMtheory}. The bosonic
string is then a stretched membrane across the interval; the orbifold breaks
translation symmetry, thus the massless vector does not appear. An orbifold
construction was also used to eliminate the $\mathbf{24}$ vector in the
Monster CFT partition function \cite{flm,harveyb}, which suggests a $D=26+1$
origin in light of bosonic M-theory \cite{bMtheory}. In fact, the original
FLM theory is bosonic string theory on $(\mathbb{R}^{24}/\Lambda _{24})/%
\mathbb{Z}_{2}$, thus it can be regarded as certain compactification of
bosonic M-theory. It is the sporadic SCFT constructions \cite%
{harveyb,sporadicft,witten3d,dunscft} with $SO_{24}$ spinors and twisted
sector states that require a generalization of bosonic M-theory with
fermions.

Bosonic M-theory contains a three-form gauge field $C^{(3)}$ for its
M2-brane, which, if one of its indices is reduced along the compact
direction, becomes the familiar two-form $B^{(2)}$ of bosonic string theory
\cite{bMtheory}. If all components of $C^{(3)}$ are evaluated in the 26
dimensions, a (massless) 3-form ($\mathbf{2,024}$) results. In the present
work we will consider the case in which the $D=25+1$ massless 1-form ($%
\mathbf{24}$) and 3-form ($\mathbf{2,024}$) persist, and actually they give
rise to the so-called $\mathbf{\lambda }$-triality, which is the
generalization of $SO_{8}$-triality up to $SO_{24}$ in a \textquotedblleft
weaker\textquotedblright , namely reducible, way\footnote{%
\textquotedblleft Weak\textquotedblright\ triality was suggested by Eric
Weinstein in 2016, at Advances in Quantum Gravity conference (San Francisco).%
}, of the form

\begin{equation}
\underset{\wedge ^{1}}{\mathbf{24}}\oplus \underset{\wedge ^{3}}{\mathbf{%
2,024},}\quad \underset{\mathbf{\lambda }}{\mathbf{2048}},\quad \underset{%
\mathbf{\lambda }_{c}}{\mathbf{2,048}^{\prime }},
\end{equation}%
\noindent relating three $2,048$-dimensional representations of $SO_{24}$
(with the subscript \textquotedblleft $c$\textquotedblright\ denoting spinor
conjugation).

\subsection{Lifting the M2-brane to $D=26+1$ and the Leech lattice}

In $D=10+1$, the presence of the M2-brane breaks the Poincar\'{e} symmetry
from $SO_{10,1}$ to $SO_{2,1}\otimes SO_{8}$, with $SO_{8}$ being the $%
\mathcal{R}$-symmetry. The near-horizon geometry of the M2-brane is given by
$AdS_{4}\otimes S^{7}$. From $D=26+1$ bosonic M-theory, the Poincar\'{e}
symmetry breaks from $SO_{26,1}$ to $SO_{2,1}\otimes SO_{24}$, where the $%
\mathcal{R}$-symmetry gets enhanced to $SO_{24}$, and the near-horizon
geometry is $AdS_{4}\otimes S^{23}$ \cite{bMtheory}. By dimensional
reduction, one can obtain a $AdS_{3}\otimes S^{23}$ background (i.e., a
generalized black string geometry), and make contact with Witten's
three-dimensional BTZ black hole \cite{witten3d}, by noting that the Conway
group $Co_{0}$, the automorphism group of the \textit{Leech lattice} $%
\Lambda _{24}$, is a maximal finite subgroup of $SO_{24}$. Geometrically,
the $196,560$ norm four Leech vectors\footnote{%
It is interesting to observe that $196,560$ is not the dimension of a unique
irrepr. of $Co_{0}$, but rather it can be decomposed as a sum of dimensions
of irreprs. of $Co_{0}$ \cite{Reprs-Co_0}. Remarkably, such a decomposition
can be made purely in terms of irreprs. of $SO_{24}$ which all survive (and
stay irreducible) under the maximal reduction $SO_{24}\rightarrow Co_{0}$,
namely : $196,560=2\cdot \mathbf{95,680}\oplus \mathbf{4,576}\oplus 2\cdot
\mathbf{276}\oplus 3\cdot \mathbf{24}$, which, by constraining the
cardinality of $\mathbf{299}$ (massless graviton in $25+1$) not to exceed $1$
(we will do this throughout the whole present paper), can also be rewritten
as $196,560=2\cdot \mathbf{95,680}\oplus \mathbf{4,576}\oplus \mathbf{299}%
\oplus \mathbf{276}\oplus 2\cdot \mathbf{24}\oplus \mathbf{1}$.} form a
discrete $S^{23}$ with symmetry given by the Conway group $Co_{0}$ \cite%
{conwaybook,wilsonLeech}. The quotient $Co_{0}/\mathbb{Z}_{2}$ yields the
simple Conway group $Co_{1}$ \cite{conwaybook}, where $2^{1+24}.Co_{1}$ is a
maximal subgroup of the Monster group $\mathbb{M}$ itself.

The set of norm four (i.e., minimal) Leech vectors is composed of three
types of elements \cite{flm}:

\begin{eqnarray}
\Lambda _{4} &=&\Lambda _{4}^{1}\cup \Lambda _{4}^{2}\cup \Lambda _{4}^{3},
\\
|\Lambda _{4}| &=&|\Lambda _{4}^{1}|+|\Lambda _{4}^{2}|+|\Lambda
_{4}^{3}|=97,152+276\cdot 4+98,304=196,560.
\end{eqnarray}%
Later, we show how to naturally recover these three types of elements from
the field content of a Monstrous M-theory in $D=26+1$. Moreover, the
assignment permits a $\mathbb{Z}_{2}$ identification, that reduces $196,560$
to $98,280$, which can occur via an orbifold. Therefore, by carefully
mapping $D=25+1$ fields descending from Monstrous M-theory to the three
types of norm four Leech vectors, and assigning the remaining fields to the
degree two piece of the Monster VOA (the Griess algebra), the construction
of the Moonshine module by FLM allows an action of the Monster $\mathbb{M}$
\cite{flm}.

\subsection{Superalgebras and central extensions}

\subsubsection{From $10+1$...}

Recalling some off-shell$~SO_{10,1}$ representations and their Dynkin labels%
\footnote{%
In an odd number of dimensions (i.e. for $\mathfrak{b}_{n}$), the rank-2
symmetric bi-spinor is equivalent to the $n$-form representation (if this is
interpreted as an $n$-brane, its Hodge dual is the $\left( n-3\right) $%
-brane). In $10+1$ dimensions $n=5$, whereas in $26+1$ dimensions $n=13$.} :%
\begin{equation}
\begin{array}{cc}
\underset{\wedge ^{1}}{\mathbf{11}}\mathbf{:} & \left( 1,0^{4}\right) ; \\
\underset{\text{(0-form)~spinor }\mathbf{\lambda }}{\mathbf{32}}\mathbf{:} &
\left( 0^{4},1\right) ; \\
\underset{\wedge ^{2}}{\mathbf{55}}\mathbf{:} & \left( 0,1,0^{3}\right) ; \\
\underset{\text{rank-2}~\text{symm.~on~spinor}}{\mathbf{462}}\simeq \underset%
{\wedge ^{5}}{\binom{11}{5}}: & \left( 0^{4},2\right) ,%
\end{array}%
\end{equation}%
the central charges that extend the $\mathcal{N}=1$, $D=10+1$ superalgebra
(i.e., the M-theory superalgebra) can be computed from the anticommutator of
the $\mathbf{2}^{\frac{11-1}{2}}\equiv \mathbf{32}$ Majorana spinor
supercharge,%
\begin{eqnarray}
\underset{32\cdot 33/2=528}{\mathbf{32}\otimes _{s}\mathbf{32}}~ &=&~%
\underset{\text{1-form }P_{\mu }}{\mathbf{11}}\oplus \underset{\text{M2}}{%
\mathbf{55}}\oplus \underset{\text{M5}}{\mathbf{462}}, \\
\text{with~Hodge~duality} &:&\underset{\text{M2}}{2}\rightarrow 4\rightarrow
11-4=7\rightarrow \underset{\text{M5}}{5},
\end{eqnarray}%
thus yielding (here $\alpha ,\beta =1,...,32$, whereas $\mu $-indices run $%
0,1,...,10$)%
\begin{equation}
\left\{ Q_{\alpha },Q_{\beta }\right\} =\left( \Gamma ^{\mu }C^{-1}\right)
_{\alpha \beta }P_{\mu }+\frac{1}{2}\left( \Gamma ^{\mu _{1}\mu
_{2}}C^{-1}\right) _{\alpha \beta }\underset{\text{M2}}{Z_{\left[ \mu
_{1}\mu _{2}\right] }^{(2)}}+\frac{1}{5!}\left( \Gamma ^{\mu _{1}...\mu
_{5}}C^{-1}\right) _{\alpha \beta }\underset{\text{M5}}{Z_{\left[ \mu
_{1}...\mu _{5}\right] }^{(5)}}.  \label{central-11}
\end{equation}

The M-theory superalgebra has an higher dimensional origin. In fact, the
central extensions in $D=10+1$ in the r.h.s. of (\ref{central-11}) can be
obtained by a Kaluza-Klein timelike-reduction of the $(1,0)$ minimal chiral
superalgebra in $D=10+2$, whose central extensions read (cfr. (3.6) of \cite%
{geoEYM} with $\mathbf{n}=0$; $\hat{\mu}$-indices here run $\mathring{0}$,$%
0,1,...,10$)%
\begin{equation}
\left\{ Q_{\alpha },Q_{\beta }\right\} =\frac{1}{2}\left( \Gamma ^{\hat{\mu}%
_{1}\hat{\mu}_{2}}C^{-1}\right) _{\alpha \beta }Z_{\left[ \hat{\mu}_{1}\hat{%
\mu}_{2}\right] }^{(2)}+\frac{1}{6!}\left( \Gamma ^{\hat{\mu}_{1}...\hat{\mu}%
_{6}}C^{-1}\right) _{\alpha \beta }Z_{\left[ \hat{\mu}_{1}...\hat{\mu}_{6}%
\right] }^{(6)}.  \label{central-12}
\end{equation}%
By splitting $\hat{\mu}=\mathring{0},\mu $, one indeed obtains%
\begin{eqnarray}
Z_{\hat{\mu}_{1}\hat{\mu}_{2}}^{(2)} &\rightarrow &\left\{
\begin{array}{l}
Z_{\mu _{1}\mathring{0}}^{(2)}\sim P_{\mu }; \\
Z_{\mu _{1}\mu _{2}}^{(2)};%
\end{array}%
\right. \\
Z_{\hat{\mu}_{1}...\hat{\mu}_{6}}^{(6)} &\rightarrow &\left\{
\begin{array}{l}
Z_{\mu _{1}...\mu _{5}\mathring{0}}^{(5)}\sim Z_{\mu _{1}...\mu _{5}}^{(5)};
\\
Z_{\mu _{1}...\mu _{6}}^{(6)}\rightarrow \epsilon _{\mu _{1}...\mu
_{11}}Z_{\nu _{6}...\nu _{11}}^{(6)}\eta ^{\mu _{6}\nu _{6}}...\eta ^{\mu
_{11}\nu _{11}}\sim Z_{\mu _{1}...\mu _{5}}^{(5)},%
\end{array}%
\right.
\end{eqnarray}%
and therefore (\ref{central-12}) yields to (\ref{central-11}).

From the r.h.s. of (\ref{central-11}), in terms of on-shell$~SO_{9}$
representations,%
\begin{equation}
\begin{array}{ccc}
\underset{g}{\mathbf{44}}\mathbf{:} & \left( 2,0^{3}\right) & \text{bosons};
\\
\underset{\text{M2~(3-form~pot. }\wedge ^{3}\text{)}}{\mathbf{84}}\mathbf{:}
& \left( 0^{2},1,0\right) & \text{bosons}; \\
\underset{\text{\textit{gravitino} (1-form spinor)~}\mathbf{\psi }}{\mathbf{%
128}}\mathbf{:} & \left( 1,0^{2},1\right) & \text{fermions},%
\end{array}%
\end{equation}%
one obtains the field content of the massless multiplet of M-theory (i.e.,
of $\mathcal{N}=1$, $D=10+1$ supergravity), having%
\begin{equation}
\underset{44+84}{B}=\underset{128}{F}.
\end{equation}

\subsubsection{...to $26+1$}

Let us generalize this to $D=(10+16)+1=26+1$ space-time dimensions. We start
some off-shell$~SO_{26,1}$ representations and their Dynkin labels,%
\begin{equation}
\begin{array}{cc}
\underset{\wedge ^{1}}{\mathbf{27}}\mathbf{:} & \left( 1,0^{12}\right) ; \\
\underset{\mathbf{\lambda }}{\mathbf{8,192}}\mathbf{:} & \left(
0^{12},1\right) ; \\
\underset{\wedge ^{2}}{\mathbf{351}}\mathbf{:} & \left( 0,1,0^{11}\right) ;
\\
\underset{\wedge ^{5}}{\mathbf{80,730}}\mathbf{:} & \left(
0^{4},1,0^{8}\right) ; \\
\underset{\wedge ^{6}}{\mathbf{296,010}}\mathbf{:} & \left(
0^{5},1,0^{7}\right) ; \\
\underset{\wedge ^{9}}{\mathbf{4,686,825}}\mathbf{:} & \left(
0^{8},1,0^{4}\right) ; \\
\underset{\wedge ^{10}}{\mathbf{8,436,285}}\mathbf{:} & \left(
0^{9},1,0^{3}\right) ; \\
\underset{\text{rank-2}~\text{symm.~on~spinor}}{\mathbf{20,058,300}}=%
\underset{\wedge ^{13}}{\binom{27}{13}}: & \left( 0^{12},2\right) .%
\end{array}%
\end{equation}%
Thus, the central charges that extends the $\mathcal{N}=1$, $D=26+1$
superalgebra can be computed from the anticommutator of the $\mathbf{2}^{%
\frac{27-1}{2}}\equiv \mathbf{8,192}$ Majorana spinor supercharge,%
\begin{eqnarray}
\underset{8,192\cdot 8,193/2=33,558,528}{\mathbf{8,192}\otimes _{s}\mathbf{%
8,192}}~ &=&~\underset{P_{\mu }}{\mathbf{27}}\oplus \underset{\text{M2}}{%
\mathbf{351}}\oplus \underset{\text{M5}}{\mathbf{80,730}}\oplus \underset{%
\text{M6}}{\mathbf{296,010}}\oplus \underset{\text{M9}}{\mathbf{4,686,825}}%
\oplus \underset{\text{M10}}{\mathbf{8,436,285}}\oplus \underset{\text{M13}}{%
\mathbf{20,058,300}};  \notag \\
&& \\
\text{with~Hodge~duality} &:&\left\{
\begin{array}{l}
\underset{\text{M2}}{2}\rightarrow 4\rightarrow 27-4=23\rightarrow \underset{%
\text{M21}}{21}; \\
\underset{\text{M5}}{5}\rightarrow 7\rightarrow 27-7=20\rightarrow \underset{%
\text{M18}}{18}; \\
\underset{\text{M6}}{6}\rightarrow 8\rightarrow 27-8=19\rightarrow \underset{%
\text{M17}}{17}; \\
\underset{\text{M9}}{9}\rightarrow 11\rightarrow 27-11=16\rightarrow
\underset{\text{M14}}{14}; \\
\underset{\text{M10}}{10}\rightarrow 12\rightarrow 27-12=15\rightarrow
\underset{\text{M13}}{13},%
\end{array}%
\right.
\end{eqnarray}%
Thus yielding (here $\alpha ,\beta =1,...,8,192$, whereas $\mu $-indices run
$0,1,...,26$)
\begin{eqnarray}
\left\{ Q_{\alpha },Q_{\beta }\right\} &=&\left( \Gamma ^{\mu }C^{-1}\right)
_{\alpha \beta }P_{\mu }+\frac{1}{2}\left( \Gamma ^{\mu _{1}\mu
_{2}}C^{-1}\right) _{\alpha \beta }\underset{\text{M2}}{Z_{\left[ \mu
_{1}\mu _{2}\right] }^{(2)}}+\frac{1}{5!}\left( \Gamma ^{\mu _{1}...\mu
_{5}}C^{-1}\right) _{\alpha \beta }\underset{\text{M5}}{Z_{\left[ \mu
_{1}...\mu _{5}\right] }^{(5)}}  \notag \\
&&+\frac{1}{6!}\left( \Gamma ^{\mu _{1}...\mu _{6}}C^{-1}\right) _{\alpha
\beta }\underset{\text{M6}}{Z_{\left[ \mu _{1}...\mu _{6}\right] }^{(6)}}+%
\frac{1}{9!}\left( \Gamma ^{\mu _{1}...\mu _{9}}C^{-1}\right) _{\alpha \beta
}\underset{\text{M9}}{Z_{\left[ \mu _{1}...\mu _{9}\right] }^{(9)}}  \notag
\\
&&+\frac{1}{10!}\left( \Gamma ^{\mu _{1}...\mu _{10}}C^{-1}\right) _{\alpha
\beta }\underset{\text{M10}}{Z_{\mu _{1}...\mu _{10}}^{(10)}}+\frac{1}{13!}%
\left( \Gamma ^{\mu _{1}...\mu _{13}}C^{-1}\right) _{\alpha \beta }Z_{%
\underset{\text{M13}}{\mu _{1}...\mu _{13}}}^{(13)}.  \label{central-27}
\end{eqnarray}

Also the $\mathcal{N}=1$, $D=26+1$ superalgebra has an higher dimensional
origin. In fact, the central extensions in $D=26+1$ in the r.h.s. of (\ref%
{central-27}) can be obtained by a Kaluza-Klein timelike-reduction from the $%
(1,0)$ minimal chiral superalgebra in $D=26+2$, whose central extensions
read (cfr. (3.6) of \cite{geoEYM} with $\mathbf{n}=2$; $\hat{\mu}$-indices
here run $\mathring{0}$,$0,1,...,26$)%
\begin{eqnarray}
\left\{ Q_{\alpha },Q_{\beta }\right\} &=&\frac{1}{2}\left( \Gamma ^{\hat{\mu%
}_{1}\hat{\mu}_{2}}C^{-1}\right) _{\alpha \beta }Z_{\left[ \hat{\mu}_{1}\hat{%
\mu}_{2}\right] }^{(2)}+\frac{1}{6!}\left( \Gamma ^{\hat{\mu}_{1}...\hat{\mu}%
_{6}}C^{-1}\right) _{\alpha \beta }Z_{\left[ \hat{\mu}_{1}...\hat{\mu}_{6}%
\right] }^{(6)}  \notag \\
&&+\frac{1}{10!}\left( \Gamma ^{\hat{\mu}_{1}...\hat{\mu}_{10}}C^{-1}\right)
_{\alpha \beta }Z_{\left[ \hat{\mu}_{1}...\hat{\mu}_{10}\right] }^{(10)}+%
\frac{1}{14!}\left( \Gamma ^{\hat{\mu}_{1}...\hat{\mu}_{14}}C^{-1}\right)
_{\alpha \beta }Z_{\left[ \hat{\mu}_{1}...\hat{\mu}_{14}\right] }^{(14)}.
\label{central-28}
\end{eqnarray}%
By splitting $\hat{\mu}=\mathring{0},\mu $, one indeed obtains%
\begin{eqnarray}
Z_{\hat{\mu}_{1}\hat{\mu}_{2}}^{(2)} &\rightarrow &\left\{
\begin{array}{l}
Z_{\mu _{1}\mathring{0}}^{(2)}\sim P_{\mu }; \\
Z_{\mu _{1}\mu _{2}}^{(2)};%
\end{array}%
\right. \\
Z_{\hat{\mu}_{1}...\hat{\mu}_{6}}^{(6)} &\rightarrow &\left\{
\begin{array}{l}
Z_{\mu _{1}...\mu _{5}\mathring{0}}^{(5)}\sim Z_{\mu _{1}...\mu _{5}}^{(5)};
\\
Z_{\mu _{1}...\mu _{6}}^{(6)};%
\end{array}%
\right. \\
Z_{\hat{\mu}_{1}...\hat{\mu}_{10}}^{(10)} &\rightarrow &\left\{
\begin{array}{l}
Z_{\mu _{1}...\mu _{9}\mathring{0}}^{(10)}\sim Z_{\mu _{1}...\mu _{9}}^{(9)};
\\
Z_{\mu _{1}...\mu _{10}}^{(10)},%
\end{array}%
\right. \\
Z_{\hat{\mu}_{1}...\hat{\mu}_{14}}^{(14)} &\rightarrow &\left\{
\begin{array}{l}
Z_{\mu _{1}...\mu _{13}\mathring{0}}^{(14)}\sim Z_{\mu _{1}...\mu
_{13}}^{(13)}; \\
Z_{\mu _{1}...\mu _{14}}^{(14)}\rightarrow \epsilon _{\mu _{1}...\mu
_{27}}Z_{\nu _{14}...\nu _{27}}^{(14)}\eta ^{\mu _{14}\nu _{14}}...\eta
^{\mu _{27}\nu _{27}}\sim Z_{\mu _{1}...\mu _{13}}^{(13)},%
\end{array}%
\right.
\end{eqnarray}%
and therefore (\ref{central-28}) yields to (\ref{central-27}).

We will elaborate on the possible existence of local supersymmetry in $26+1$
further below. For the time being, we confine ourselves to observe that
Susskind and Horowitz identified a subset of above (central, $p$-brane)
charges for bosonic M-theory \cite{bMtheory}, whereas the most general set
of central extensions is provided by the r.h.s. of (\ref{central-27}). We
note that the automorphic form of the fake Monster Lie algebra satisfies
functional equations generated by transformations in the group Aut$%
(II_{26,2})^{+}$ \cite{boraut}, a discrete subgroup of $O_{26,2}$ which can
transform fields in $D=26+2$, $D=26+1$ and $D=25+1$. Thus, the signature $%
D=26+2$ has proven essential in the proof of Monstrous Moonshine, and it
gives further evidence for an M-theoretical origin. We can anticipate that
Monstrous M-theory, in fact, has its most natural origin in $D=29+1$ or $%
D=28+2$ with purely bosonic massless states descending from 5-form and dual
23-form gauge fields, respectively of a 4-brane and 22-brane. By dimensional
reduction to $D=26+1$, such higher 5-form and 23-form gauge fields break up
non-trivially, providing a rich structure to possibly realize a would-be
supergravity with a $\mathbf{98,304}$ Rarita-Schwinger field, as we will see
in Sec. \ref{BBFF}.

\subsection{M-branes, Horava-Witten and the Monster SCFT}

The presence of an M10-brane breaks $SO_{26,1}$ Poincar\'{e} symmetry to $%
SO_{10,1}\otimes SO_{16}$, giving $D=10+1$ Poincar\'{e} symmetry on its
worldvolume \cite{wmtheory}. The $\mathbf{8,192}$ spinor then factorizes as $%
(\mathbf{32},\mathbf{128})\oplus (\mathbf{32},\mathbf{128}^{\prime })$, thus
isolating a hidden $\mathbf{128}^{(\prime )}$ spinor, which can be used to
form $\mathfrak{e}_{8}=\mathfrak{so}_{16}\oplus \mathbf{128}^{(\prime )}$.
Intriguingly, this may suggest an origin for Horava-Witten theory \cite%
{hwtheory1,hwtheory2},which requires an eleven-manifold $M^{11}$ with
boundary, whose boundary points are the $\mathbb{Z}_{2}$ fixed points in $%
M^{11}$; in this theory, a M2-brane stretched between these fixed points
yields the strongly coupled heterotic string \cite{hwtheory1,hwtheory2}. On
the other hand, in the presence of the broken $\mathbf{8,192}$ spinor we see
a possible reason for the $E_{8}$ symmetry that arises at the fixed points,
as the hidden spinor fermions may contribute to anomalies induced via
orbifold of the M10-brane worldvolume.

If instead we reduce directly from $D=26+1$ on an orbifold $S^{1}/\mathbb{Z}%
_{2}$, we break half the supersymmetry and remove the $\mathbf{24}$ vector,
while the $\mathbf{8,192}$ spinor projects down to $\mathbf{4,096}$. This is
in agreement with the Monster SCFT \cite{harveyb}, where the fixed points of
the orbifold contain $4,096$ twisted states. This differs from the orbifold
reduction of bosonic M-theory, where the fixed points have no extra degrees
of freedom due to the absence of chiral bosons and fermions \cite{bMtheory}.
A $D=26+1$ M-theory with $\mathbf{24}\cdot\mathbf{4,096}=\mathbf{98,304}$
Rarita-Schwinger field would have fermionic anomalies at each orbifold fixed
point, that must be canceled by vector multiplets as in the $D=10+1$
M-theory case \cite{hwtheory1,hwtheory2}. One would expect a generalization
of $E_8$ symmetry at each fixed point, that contains at least $\mathbf{24}%
\cdot\mathbf{2^{12}}=\mathbf{98,304}$ vector multiplets for RNS twisted
sector states. The Griess algebra provides such minimal degrees of freedom,
thus could possibly be used to cancel anomalies at the fixed points. Another
possibility is the \textit{Leech algebra}, which we introduce in a later
section.

It was shown that bosonic M-theory can reduce to the bosonic string in $%
D=25+1$ by reduction along $S^{1}/\mathbb{Z}_{2}$ \cite{bMtheory}, thus the
Monster CFT in its relation to $D=25+1$ bosonic string theory on $(\mathbb{R}%
^{24}/\Lambda _{24})/\mathbb{Z}_{2}$ can trace its origin back to $26+1$
space-time dimensions. This suggests that the Monster CFT describes states
on the boundary of $AdS_{3}\otimes S^{23}$, originating from the M2-brane
near-horizon geometry $AdS_{4}\otimes S^{23}$, where the transverse
directions are discretized and given by the Leech lattice $\Lambda _{24}$.
In going from $D=26+1$ to $D=25+1 $ the $\mathbf{324}$ graviton breaks as $%
\mathbf{324}=\mathbf{299}\oplus \mathbf{24}\oplus \mathbf{1}$, where the
orbifold removes translation symmetry, and hence eliminates the 1-form $%
\mathbf{24}$ \cite{bMtheory} from the closed string spectrum \cite{bMtheory}.

Recall, that a holomorphic CFT for the Leech lattice $\Lambda _{24}$ has
partition function
\begin{eqnarray}
Z_{\text{Leech}}(q) &=&\frac{\Theta _{\Lambda _{24}}}{\eta ^{24}}=\frac{1}{q}%
+24+196,884q+21,493,760q^{2}+864,299,970q^{3}+\mathcal{O}(q^{4})
\label{Z-Leech} \\
&=&J(q)+24,
\end{eqnarray}%
where $J(q)=j(q)-744$, and $j(q)$ is the $j$\textit{-function} \cite{harveyb}%
. FLM have used a $\mathbb{Z}_{2}$-twisted version of the Leech theory to
remove the unwanted $\mathbf{24}$ states that contribute to the constant
term in the partition function and to obtain the appropriate finite group
structure \cite{flm,harveyb}. From a modern perspective, this can be
accomplished by a $S^{1}/\mathbb{Z}_{2}$ orbifold reduction of $D=26+1$
bosonic M-theory, which has been shown to reduce to the $D=25+1$ bosonic
string \cite{bMtheory}.

A superconformal field theory (\textquotedblleft the Beauty and the
Beast\textquotedblright ) description of the FLM model was given by Dixon,
Ginsparg and Harvey \cite{harveyb}. The supersymmetric extension of the
Virasoro algebra introduces moments $G_{n}$ that satisfy the relations

\begin{equation}
[L_m,G_n]=\left(\frac{m}{2}-n\right)G_{m+n}
\end{equation}
and

\begin{equation}
\left\{ G_{m},G_{n}\right\} =2L_{m+n}+\frac{\hat{c}}{2}(m^{2}-\frac{1}{4}%
)\delta _{m+n,0}.
\end{equation}%
For integer moding of $G_{n}$ ($n\in \mathbb{Z}$), the supersymmetric
extension is named the Ramond (R) algebra, while for half-integer moding ($%
n\in \mathbb{Z}+\frac{1}{2}$) it is named the Neveu-Schwarz (NS) algebra
\cite{harveyb}. The $\mathbf{2}^{12}\equiv \mathbf{4,096}$ twisted states of
the FLM model are half-integer moded \cite{flm,harveyb}, thus suggesting a
fermionic origin. This can arise from projecting half the degrees of freedom
of the $\mathbf{8,192}$ spinor from $D=26+1$, which is expected from an
orbifold reduction, analogous to the case of $D=10+1$ M-theory where the $%
\mathbf{32}$ spinor is projected to a $\mathbf{16}$ \cite%
{hwtheory1,hwtheory2}. A $S^{1}/\mathbb{Z}_{2}$ orbifold reduction reduces
the $\mathbf{8,192}$ spinor to $\mathbf{4,096}$ spinor, where under $SO_{24}$
one has $\mathbf{4,096}=\mathbf{2,048}\oplus \mathbf{2,048}^{\prime }$. The $%
\mathbf{2048}$ spinors can yield worldsheet fermions in $D=25+1$. Such $%
SO_{24}$ spinors are seen in Duncan's SCFT with Conway group symmetry \cite%
{dunscft}. These spinors can be used to build RNS states in $D=25+1$, that
generalize the gravitino and dilatino states of type IIA in $D=9+1$ from $%
128 $ to $98,304$ degrees of freedom.

The untwisted states of the FLM model include the $24$ Ramond ground states
and the $196,560/2=98,280$ Leech lattice states, which $G_{0}$ pairs with $%
24\cdot 2^{12}=98,304$ dimension 2 Ramond fields as $98,280+24=98,304$ \cite%
{flm,harveyb}. In $D=26+1$, the massless Rarita-Schwinger (1-form spinor)
field has $\mathbf{98,304}$ degrees of freedom, thus is a candidate for the
origin of the dimension 2 Ramond fields in a SCFT. The remaining $98,280$
states come from a discretized transverse space, where in the $%
AdS_{4}\otimes S^{23}$ near-horizon geometry of the M2-brane in $D=26+1$ the
23-sphere is discretized by the $196,560$ norm four Leech lattice vectors.
This is consistent with the Conway group $Co_{0}$ being a maximal finite
subgroup of the $\mathcal{R}$-symmetry group $SO_{24}$. The $S^{1}/\mathbb{Z}%
_{2}$ orbifold reduces the $196,560$ vectors to $98,280$, while also
reducing $AdS_{4}\otimes S^{23}$ to $AdS_{3}\otimes S^{23}$, and breaking
the discrete $\mathcal{R}$-symmetry group $Co_{0}$ down to the simple Conway
group\footnote{%
It is interesting to observe that $\mathbf{98,280}$ is not the dimension of
a unique irrepr. of $Co_{1}$, but rather it can be decomposed as a sum of
irreprs. of $Co_{1}$ \cite{Reprs-Co_1}. Remarkably, one finds a
decomposition only in terms of irreprs. of $Co_{0}$ which all survive (and
stay irreducible) under the maximal reduction $Co_{0}\rightarrow Co_{1}$,
namely $\mathbf{98,280}=\mathbf{80,730}\oplus \mathbf{17,250}\oplus \mathbf{%
299}\oplus \mathbf{1}$.} $Co_{1}\simeq Co_{0}/\mathbb{Z}_{2}$ \cite%
{conwaybook}, thus making contact with Witten's holographic interpretation
of the Monster \cite{witten3d} with $Co_{1}$ as a discrete R-symmetry.

Finally, it is here worth mentioning that the $\mathbb{Z}_{2A}$%
-fermionization of the Monster CFT \cite{fmonster} reveals representations
of the Baby Monster finite group $\mathbb{BM}$ in the NS and R sectors,

\begin{equation}
Z_{NS}^{\mathcal{F}}(\tau )=\frac{1}{q}+\frac{1}{\sqrt{q}}+4,372\sqrt{q}%
+100,628q+\mathcal{O}(q^{3/2}),
\end{equation}

and

\begin{equation}
Z_{R}^{\mathcal{F}}(\tau )=192,512q+21,397,504q^{2}+\mathcal{O}(q^{3}),
\end{equation}%
where $4,372+192,512=196,884$. In terms of $SO_{24}$ irreprs., we note that $%
4,372=\mathbf{276}\oplus \mathbf{2,048}\oplus \mathbf{2,048}^{\prime }$,
where worldsheet fermions are suggested. This implies a $D=25+1$ string
theory with $\mathbf{2,048}\oplus \mathbf{2,048}^{\prime }$ worldsheet
fermions that generalizes the $D=9+1$ superstring with $SO_{8}$ spinors. In
the treatment given below, we propose a $D=26+1$ origin for such a string
theory, supported by the fermionization of the Monster CFT \cite{fmonster},
which suggests a $(2+1)$-dimensional fermionic gravitational Chern-Simons
term that can live on the boundary of $AdS_{4}$. Once again, given an $S^{1}/%
\mathbb{Z}_{2}$ orbifold reduction of $D=26+1$ M-theory with fermions, one
does expect anomalies, and to cancel such anomalies may necessitate the use
of the Leech lattice $\Lambda _{24}$ at each fixed hyperplane. The resulting
$D=25+1$ closed string theory is then very similar to the Bimonster string
theory introduced by Harvey et al. in \cite{bimonster}.

\section{\label{triality}\textquotedblleft Weak\textquotedblright\
trialities in 24 dimensions}

By \textit{triality}, denoted by $\mathbb{T}$, in this paper we refer to a
property of the Lie algebra $\mathfrak{d}_{4}$ (see e.g. \cite{triality}),
namely a map among its three $8$-dimensional irreducible representations%
\begin{equation}
\mathfrak{d}_{4}:\left\{
\begin{array}{l}
\mathbf{\wedge }^{1}\equiv \mathbf{8}_{v}:=(1,0,0,0)~\left( \text{1-form}%
\right) ; \\
~ \\
\mathbf{\lambda }\equiv \mathbf{8}_{s}:=(0,0,0,1)~\left( \text{semispinor}%
\right) ; \\
~ \\
\mathbf{\lambda }^{\prime }\equiv \mathbf{\lambda }_{c}\equiv \mathbf{8}%
_{s}^{\prime }\equiv \mathbf{8}_{c}:=(0,0,1,0)~\left( \text{%
conjugate~semispinor}\right)%
\end{array}%
\right.
\end{equation}%
among themselves :%
\begin{equation}
\mathbb{T}:%
\begin{array}{ccc}
\mathbf{\wedge }^{1} & ~ & ~ \\
\uparrow \downarrow & \searrow \nwarrow & ~ \\
\mathbf{\lambda }^{\prime } & \rightleftarrows & \mathbf{\lambda }%
\end{array}%
.  \label{T}
\end{equation}%
The origin of $\mathbb{T}$ can be traced back to the three-fold structural
symmetry of the Dynkin diagram of $\mathfrak{d}_{4}$, and to the existence
of an outer automorphism of $\mathfrak{d}_{4}$ which interchanges $\mathbf{8}%
_{v}$, $\mathbf{8}_{s}$ and $\mathbf{8}_{c}$; in fact, the outer
automorphism group of $\mathfrak{d}_{4}$ (or, more precisely, of the
corresponding spin group $Spin_{8}$, the double cover of the Lie group $%
SO_{8}$) is isomorphic to the symmetric group $S_{3}$ that permutes such
three representations.

Thence, through suitably iterated tensor products of representations $%
\mathbf{8}_{v}$, $\mathbf{8}_{s}$ and $\mathbf{8}_{c}$, $\mathbb{T}$ affects
higher-dimensional representations, as well. For instance, $\mathbb{T}$ maps
also the three $56$-dimensional irreducible representations of $\mathfrak{d}%
_{4}$ :%
\begin{equation}
\mathfrak{d}_{4}:\left\{
\begin{array}{l}
\mathbf{\wedge }^{3}\equiv \mathbf{56}_{v}:=(0,0,1,1)~\left( \text{3-form }%
\wedge ^{3}\right) ; \\
~ \\
\mathbf{\psi }\equiv \mathbf{56}_{s}:=(1,0,0,1)~\left( \text{1-form spinor,
\textit{aka} gravitino}\right) ; \\
~ \\
\mathbf{\psi }_{c}\equiv \mathbf{\psi }^{\prime }\equiv \mathbf{56}%
_{s}^{\prime }\equiv \mathbf{56}_{c}:=(1,0,1,0)~\left( \text{%
conjugate~gravitino}\right)%
\end{array}%
\right.
\end{equation}%
among themselves%
\begin{equation}
\mathbb{T}:%
\begin{array}{ccc}
\mathbf{\wedge }^{3} & ~ & ~ \\
\uparrow \downarrow & \searrow \nwarrow & ~ \\
\mathbf{\psi } & \rightleftarrows & \mathbf{\psi }^{\prime }%
\end{array}%
.  \label{T'}
\end{equation}%
By \textit{gravitino}, we mean the gamma-traceless 1-form spinor; indeed, in
order to correspond to an irreducible representation, the spinor-vector $%
\psi _{\mu }^{\alpha }$ must be gamma-traceless :%
\begin{equation}
\Gamma _{\alpha \beta }^{\mu }\psi _{\mu }^{\beta }=0,
\end{equation}%
where $\mu $ and $\alpha $ are the vector resp. spinor indices, and $\Gamma
_{\alpha \beta }^{\mu }\equiv \left( \Gamma ^{\mu }\right) _{\alpha \beta }$
denote the gamma matrices of $\mathfrak{d}_{4}$. $\psi $ is a
Rarita-Schwinger (RS) field of spin/helicity $\frac{3}{2}$, and, in the
context of supersymmetric theories, it is named \textit{gravitino }(being
the spartner of the graviton $g_{\mu \nu }$). As (\ref{T}) denotes the
action of triality $\mathbb{T}$ on (semi)spinors, (\ref{T'}) expresses the
triality $\mathbb{T}$ acting on RS fields. $\mathbb{T}$ plays an important
role in type II string theory in $9+1$ space-time dimensions, in which $%
\mathfrak{so}_{8}$ (compact real form of $\mathfrak{d}_{4}$) is the algebra
of the massless little group (cfr. e.g. \cite{triality-string}).

\subsection{\label{lambda}$\mathbf{\protect\lambda }$-triality}

In certain dimensions, there may be a \textquotedblleft weaker" variant of $%
\mathbb{T}$, in which $\mathbf{\lambda }$ and $\mathbf{\lambda }^{\prime }$
have the same dimension of a \textit{reducible} (bosonic) representation,
namely of a sum of \textit{irreducible} (bosonic) representations, of $%
\mathfrak{d}_{n}$. In fact, for $n=12$ (i.e. in $\mathfrak{d}_{12}$)
something remarkable takes place: in $\mathfrak{d}_{12}$, the following
three representations all have the same dimension $2,048$:%
\begin{equation}
\mathfrak{d}_{12}:\left\{
\begin{array}{l}
\mathbf{\lambda }\equiv \mathbf{2}^{11}=\mathbf{2,048}:=(0^{11},1); \\
~ \\
\mathbf{\lambda }^{\prime }\equiv \left( \mathbf{2}^{11}\right) ^{\prime }=%
\mathbf{2,048}^{\prime }:=(0^{10},1,0); \\
~ \\
\mathbf{\wedge }^{1}\oplus \mathbf{\wedge }^{3}=\mathbf{24}\oplus \mathbf{%
2,024}=(1,0^{11})\oplus (0^{2},1,0^{9}).%
\end{array}%
\right.  \label{1}
\end{equation}%
In other words, in $\mathfrak{d}_{12}$ the \textit{reducible} bosonic
representation given by the sum of the vector (1-form) representation $%
\mathbf{\wedge }^{1}$ and of the 3-form representation $\mathbf{\wedge }^{3}$
has the same dimension of each of the (semi)spinors $\mathbf{\lambda }$ and $%
\mathbf{\lambda }^{\prime }$. Analogously to the aforementioned case of $%
\mathfrak{d}_{4}$, one can then define a \textquotedblleft
triality-like\textquotedblright\ map, named $\mathbf{\lambda }$-\textit{%
triality} and denoted by $\mathbb{\tilde{T}}_{\mathbf{\lambda }}$, between
the corresponding representation vector representation spaces,
\begin{equation}
\mathbb{\tilde{T}}_{\mathbf{\lambda }}:%
\begin{array}{ccc}
\left( \mathbf{\wedge }^{1}\oplus \mathbf{\wedge }^{3}\right) & ~ & ~ \\
\uparrow \downarrow & \searrow \nwarrow & ~ \\
\mathbf{\lambda } & \rightleftarrows & \mathbf{\lambda }^{\prime }%
\end{array}
\label{1'}
\end{equation}%
It is immediate to realize that a crucial difference with (\ref{T}) relies
in the \textit{reducibility} of the bosonic sector of the map, which we will
henceforth associate to the \textquotedblleft weakness\textquotedblright\ of
$\mathbb{\tilde{T}}_{\mathbf{\lambda }}$. However, since no other Dynkin
diagram (besides $\mathfrak{d}_{4}$) has an automorphism group of order
greater than $2$, one can also conclude that (\ref{1})-(\ref{1'}) cannot be
realized as an automorphism of $\mathfrak{d}_{12}$, nor it can be traced
back to some structural symmetry of the Dynkin diagram of $\mathfrak{d}_{12}$
itself.

\subsection{\label{psi}$\mathbf{\protect\psi }$-triality}

As triality $\mathbb{T}$ of $\mathfrak{d}_{4}$ (\ref{T}) affects all tensor
products stemming from $\mathbf{8}_{v}$, $\mathbf{8}_{s}$ and $\mathbf{8}%
_{c} $, implying in particular (\ref{T'}), so the $\mathbf{\lambda }$%
-triality $\mathbb{\tilde{T}}_{\mathbf{\lambda }}$ of $\mathfrak{d}_{12}$ (%
\ref{1'}) affects all tensor products stemming from $\mathbf{\wedge }%
^{1}\oplus \mathbf{\wedge }^{3}$, $\mathbf{\lambda }$ and $\mathbf{\lambda }%
^{\prime }$; in particular, in $\mathfrak{d}_{12}$, the following three
representations have the same dimension $47,104$:%
\begin{equation}
\mathfrak{d}_{12}:\left\{
\begin{array}{l}
\mathbf{\psi }\equiv \mathbf{47,104}:=(1,0^{10},1); \\
~ \\
\mathbf{\psi }^{\prime }\equiv \mathbf{47,104}^{\prime }:=(1,0^{9},1,0); \\
~ \\
2\cdot \left( 2\cdot \mathbf{\wedge }^{4}\oplus \mathbf{\wedge }^{3}\oplus
\mathbf{\wedge }^{2}\right) =2\cdot \left( 2\cdot \mathbf{10,626}\oplus
\mathbf{2,024}\oplus \mathbf{276}\right) \\
=2\cdot \left( 2\cdot (0^{3},1,0^{8})\oplus (0^{2},1,0^{9})\oplus
(0,1,0^{10})\right) .%
\end{array}%
\right.  \label{2}
\end{equation}%
In other words, in $\mathfrak{d}_{12}$ the \textit{reducible} bosonic
representation given by the sum of the 4-form $\mathbf{\wedge }^{4}$, 3-form
$\mathbf{\wedge }^{3}$ and 2-form $\mathbf{\wedge }^{2}$ representations
(with multiplicity 4, 2 and 2, respectively) has the same dimension of each
of the RS field representations $\mathbf{\psi }$ and $\mathbf{\psi }^{\prime
}$. Analogously to the aforementioned case of $\mathfrak{d}_{4}$, one can
then define a \textquotedblleft triality-like" map, named $\mathbf{\psi }$%
\textit{-triality} and denoted by $\mathbb{\tilde{T}}_{\mathbf{\psi }}$,
between the corresponding representation vector spaces,
\begin{equation}
\mathbb{\tilde{T}}_{\mathbf{\psi }}:%
\begin{array}{ccc}
2\cdot \left( 2\cdot \mathbf{\wedge }^{4}\oplus \mathbf{\wedge }^{3}\oplus
\mathbf{\wedge }^{2}\right) & ~ & ~ \\
\uparrow \downarrow & \searrow \nwarrow & ~ \\
\mathbf{\psi } & \rightleftarrows & \mathbf{\psi }^{\prime }%
\end{array}%
.  \label{2'}
\end{equation}%
Again, (\ref{1})-(\ref{1'}) cannot be realized as an automorphism of $%
\mathfrak{d}_{12}$, nor it can be traced back to some structural symmetry of
the Dynkin diagram of $\mathfrak{d}_{12}$ itself.

\subsection{\label{bos}Iso-dimensionality among (sets of) $p$-forms : an
example}

Representations with the same dimensions can also be only bosonic. Still, $%
\mathfrak{d}_{12}$ provides the following example of such a phenomenon%
\footnote{%
Another example is provided by the iso-dimensionality map $\mathbf{\wedge }%
^{1}\oplus \wedge ^{2}\leftrightarrow S_{0}^{2}\oplus \mathbf{1}$, holding
for any orthogonal Lie algebra. However, since we will fix the number of
graviton fields to be $1$, we will not make use of such an
iso-dimensionality map.} : the following two sets of representations have
the same dimension $42,504$,%
\begin{equation}
\mathfrak{d}_{12}:\left\{
\begin{array}{l}
\wedge ^{5}\equiv \mathbf{42,504}:=(0^{4},1,0^{7}); \\
~ \\
4\cdot \mathbf{\wedge }^{4}=4\cdot \mathbf{10,626}=4\cdot (0^{3},1,0^{8}).%
\end{array}%
\right.  \label{3}
\end{equation}%
In other words, in $\mathfrak{d}_{12}$ the 5-form representation $\wedge
^{5} $ has the same dimension, namely $42,504$, of four copies of the 4-form
representation $\wedge ^{4}$. Again, one can then define a map, denoted by $%
\mathcal{B}$, between the corresponding representation vector spaces%
\footnote{%
Of course, all instances of iso-dimensionality among representations given
by (\ref{1})-(\ref{1'}), (\ref{2})-(\ref{2'}) and (\ref{3})-(\ref{3''}),
hold up to Poincar\'{e}/Hodge duality (in the bosonic sector); cfr. (\ref%
{poss1}) further below. Note that other iso-dimensionality maps besides (\ref%
{3''}) may exist, but we will not make use of them in the present paper.},%
\begin{equation}
\mathcal{B}:\mathbf{\wedge }^{5}\leftrightarrow 4\cdot \mathbf{\wedge }^{4}.
\label{3''}
\end{equation}


\section{\label{MGra}Monstrous Dilatonic Gravity in $25+1$}

In the previous Sec. \ref{triality} we have introduced some maps among
fermionic and bosonic representations of $\mathfrak{d}_{12}$, having the
same dimension but different Dynkin labels:

\begin{itemize}
\item the $\mathbf{\lambda }$-triality $\mathbb{\tilde{T}}_{\mathbf{\lambda }%
}$ (\ref{1})-(\ref{1'}), generalizing the triality $\mathbb{T}$ (\ref{T}) of
$\mathfrak{d}_{4}$ to $\mathfrak{d}_{12}$;

\item the $\mathbf{\psi }$-triality $\mathbb{\tilde{T}}_{\mathbf{\psi }}$ (%
\ref{2})-(\ref{2'}), extending the weak triality of $\mathfrak{d}_{12}$ to
its Rarita-Schwinger sector;

\item the iso-dimensionality map $\mathcal{B}$ (\ref{3})-(\ref{3''}) among
certain sets of bosonic ($p$-form) representations of $\mathfrak{d}_{12}$.
\end{itemize}

As triality $\mathbb{T}$ (\ref{T}) of $\mathfrak{d}_{4}$ plays a role in the
type II string theories (which all have $\mathfrak{so}_{8}$ as the algebra
of the massless little group), one might ask whether (\ref{1'}), (\ref{2'})
and (\ref{3''}) have some relevance in relation to bosonic string theory
\cite{bMtheory}, or in relation to more general field theories in $D=25+1$
space-time dimensions, in which $\mathfrak{so}_{24}$ is the algebra of the
massless little group. Below, we will show that this is actually the case
for a quite large class of \textit{non-supersymmetric} dilatonic (Einstein)
gravity theories in $25+1$, named \textit{Monstrous} \textit{gravities},
which we are now going to introduce.

To this aim, we start and display various massless fields in $D=25+1$
space-time dimensions. As mentioned, each massless field fits into the
following irreducible representation\footnote{%
A priori, one could also consider $\wedge ^{6}\equiv \mathbf{134,596}$
(because $134,596<196,884$ - see below - ), but it actually does not enter
in any way in the treatment of this Section.} $\mathbf{R}$ of the massless
little group $SO_{24}$ (recall that $g\equiv S_{0}^{2}$ and $\phi \equiv
\mathbf{1}$ throughout):%
\begin{equation}
\begin{array}{ccc}
\text{field} & \mathbf{R} & \text{{\footnotesize Dynkin labels}} \\
g: & \mathbf{299} & \left( 2,0^{11}\right) \\
\psi : & \mathbf{47,104} & \left( 1,0^{10},1\right) \\
\psi ^{\prime }: & \mathbf{47,104}^{\prime } & (1,0^{9},1,0) \\
\wedge ^{1} & \mathbf{24} & (1,0^{11}) \\
\lambda : & \mathbf{2,048} & (0^{11},1) \\
\lambda ^{\prime }: & \mathbf{2,048}^{\prime } & (0^{10},1,0) \\
\phi : & \mathbf{1} & (0^{12}) \\
\wedge ^{5}: & \mathbf{42,504} & (0^{4},1,0^{7}) \\
\wedge ^{4}: & \mathbf{10,626} & (0^{3},1,0^{8}) \\
\wedge ^{3}: & \mathbf{2,024} & (0^{2},1,0^{9}) \\
\wedge ^{2}: & \mathbf{276} & (0,1,0^{10})%
\end{array}%
\end{equation}%
We are now going to classify field theories in $25+1$ space-time dimensions
which share the following features :

\begin{description}
\item[a] They all contain gravity (in terms of \textit{one} 26-bein, then
yielding \textit{one} metric tensor $g_{\mu \nu }$) and \textit{one} dilaton
scalar field $\phi $; thus, the Lagrangian density of their
gravito-dilatonic sector reads\footnote{%
Throughout our analysis, we rely on the conventions and treatment given in
Secs. 22 and 23 of \cite{Ortin-book}.}%
\begin{equation}
\mathcal{L}=e^{-2\phi }\left( R-4\partial _{\mu }\phi \partial ^{\mu }\phi
\right) .
\end{equation}

\item[b] The relations among all such theories are due to the $\mathbf{%
\lambda }$-triality $\mathbb{\tilde{T}}_{\mathbf{\lambda }}$ (\ref{1})-(\ref%
{1'}), the \textit{weak }$\mathbf{\psi }$-triality $\mathbb{\tilde{T}}_{%
\mathbf{\psi }}$ (\ref{2})-(\ref{2'}), as well as the bosonic map $\mathcal{B%
}$ (\ref{3})-(\ref{3''}) of $\mathfrak{so}_{24}$ (real compact form of $%
\mathfrak{d}_{12}$), which is the Lie algebra of the massless little group.

\item[c] By constraining the theories to contain \textit{only one} graviton
and \textit{only one} dilaton, the total number of degrees of freedom of the
massless spectrum must sum up to%
\begin{eqnarray}
&&1+299+47,104\cdot \left( \#\psi \right) +24\cdot \left( \#\wedge
^{1}\right) +2,048\cdot \left( \#\lambda \right)  \notag \\
&&+42,504\cdot \left( \#\wedge ^{5}\right) +10,626\cdot \left( \#\wedge
^{4}\right) +2,024\cdot \left( \#\wedge ^{3}\right) +276\cdot \left(
\#\wedge ^{2}\right)  \notag \\
&=&196,884.  \label{condd}
\end{eqnarray}
\end{description}

Consequently, the whole set of massless degrees of freedom of such theories
may be acted upon by the \textit{Monster group} $\mathbb{M}$, the largest
sporadic group, because $\mathbf{196,883}$ is the dimension of its smallest
non-trivial representation \cite{griess76}. For this reason, the
gravito-dilatonic theories under consideration will all be named \textit{%
Monstrous gravities}. They will be characterized by the following split,
\begin{equation}
196,884=\mathbf{196,883}\oplus \mathbf{1},  \label{deccc}
\end{equation}%
which is at the origin of the so-called Monstrous Moonshine \cite{conway79,
borcherds}. The dilaton $\phi $, which is a singlet of $\mathbb{M}$,
coincides with the vacuum state $\left\vert \Omega \right\rangle $ of the
chiral Monster SCFT discussed in \cite{harveyb,cMonster2,cMonster3}. Thus,
Monstrous gravities in $25+1$ space-time dimensions, and the presence of a
unique $\phi $, are intimately related to the $196,883$-dimensional
representation of $\mathbb{M}$, and thus \textit{they may provide an
explanation of the (observation who firstly ignited the) Monstrous Moonshine
in terms of (higher-dimensional, gravitational) field theory}.

In the context of Witten's three-dimensional gravity \cite{witten3d}, this
suggests that the $196,883$ primary operators that create black holes are
carrying dilatonic gravity field content. As done in \cite{witten3d}, it is
enlightening to compare the number $196,883$ of primaries with the
Bekenstein-Hawking entropy of the corresponding black hole : an exact
quantum degeneracy of $196,883$ yields an entropy of Witten's BTZ black hole
given by $\ln \left( 196,883\right) \simeq 12.19$, whereas the
Bekenstein-Hawking entropy-area formula yields to $4\pi \simeq 12.57$. Of
course, one should not expect a perfect agreement between such two
quantities, because the Bekenstein-Hawking entropy-area formula holds in the
semi-classical regime and not in the exact quantum one. As given in (\ref%
{condd}), $196,883$ comes from gauge fields (potentials), graviton, etc.,
albeit without dilaton; in this sense, \textit{the quantum entropy} $\ln
\left( 196,883\right) \simeq 12.19$ \textit{has a manifest
higher-dimensional interpretation}, since the BTZ black hole degrees of
freedom can be expressed in terms of massless degrees of freedom of fields
in $25+1$ space-time dimensions.

\subsection{\label{class}Classification}

All Monstrous gravities will be classified by using two sets of numbers :

\begin{itemize}
\item $\mathbf{s}_{1}$, a length-5 string, providing the number of
independent \textquotedblleft helicity"-$h$ massless fields, with $h=2,\frac{%
3}{2},1,\frac{1}{2},0$, respectively denoted by $g$ (graviton), $\psi $
(Rarita-Schwinger field), $\wedge ^{1}$ (1-form potential), $\lambda $
(spinor field\footnote{%
The spinor field gets named \textit{gaugino} (or \textit{dilatino}) in
presence of supersymmetry.}), and $\phi $ (dilaton); as pointed out above,
we fix $\#g=\#\phi =1$ throughout\footnote{%
Also note that any theory with $\#\wedge ^{1}\geqslant 1$ is a
Maxwell-Einstein-dilaton theory in $25+1$ space-time dimensions.} :
\begin{equation}
\mathbf{s}_{1}:=\left( \#g,\#\psi ,\#\wedge ^{1},\#\lambda ,\#\phi \right)
=\left( 1,\#\psi ,\#\wedge ^{1},\#\lambda ,1\right) ;
\end{equation}

\item $\mathbf{s}_{2}$, a length-4 string, providing the number of
independent $p$-form brane potentials, for the smallest values of $p$,
namely for $p=5,4,3,2$,%
\begin{equation}
\mathbf{s}_{2}:=\left( \#\wedge ^{5},\#\wedge ^{4},\#\wedge ^{3},\#\wedge
^{2}\right) .
\end{equation}
\end{itemize}

Before starting, we should point out that the classification below is unique
up to Poincar\'{e}/Hodge duality $\ast $ for the $p$-form potentials,
namely, for $p=1,..5$ :%
\begin{equation}
\begin{array}{ccc}
{\footnotesize p}\text{{\footnotesize -form pot.}} & \overset{\ast }{%
\longleftrightarrow } & {\footnotesize p}^{\prime }\text{{\footnotesize %
-form pot.}} \\
\wedge ^{1} &  & \wedge ^{23} \\
\wedge ^{2} &  & \wedge ^{22} \\
\wedge ^{3} &  & \wedge ^{21} \\
\wedge ^{4} &  & \wedge ^{20} \\
\wedge ^{5} &  & \wedge ^{19}%
\end{array}
\label{poss1}
\end{equation}%
as well as up to chiral/non-chiral arrangements in the fermionic sector,%
\begin{eqnarray}
&&%
\begin{array}{ccc}
\#\psi &  & \text{{\footnotesize chiral/non-chiral arr.s}} \\
2 &  & \left( 2,0\right) ,(0,2),(1,1) \\
4 &  & (4,0),(0,4),(3,1),(1,3),(2,2);%
\end{array}
\label{poss2} \\
&&%
\begin{array}{ccc}
\#\lambda &  & \text{{\footnotesize chiral/non-chiral arr.s}} \\
1 &  & (1,0),(0,1) \\
2 &  & (2,0),(0,2),(1,1) \\
3 &  & (3,0),(0,3),(2,1),(1,2).%
\end{array}
\label{poss3}
\end{eqnarray}%
Clearly, (\ref{poss1}) and (\ref{poss2})-(\ref{poss3}) are particularly
relevant if (local) supersymmetry in $25+1$ were considered; however, in
this paper we will not be dealing with such an interesting topic, and we
will confine ourselves to make some comments further below (in $26+1$).

We will split the Monstrous gravity theories, sharing the features \textbf{a
}- \textbf{c} listed above, in five groups, labelled with Latin numbers : $0$%
, $1$, $2$, $3$, $4$, specifying the number $\#\psi $ of $h=3/2$ RS fields.
The $\mathbf{\psi }$-triality $\mathbb{\tilde{T}}_{\mathbf{\psi }}$ (\ref{2}%
)-(\ref{2'}) of $\mathfrak{so}_{24}$ maps such five groups among themselves.
Then, each of these groups will be split into four subgroups, labelled with
Greek letters : $\alpha $, $\beta $, $\gamma $ and $\delta $, respectively
characterized by the following values of $\#\wedge ^{1}$ and $\#\lambda $ :%
\begin{equation}
\left( \#\wedge ^{1},\#\lambda \right) =%
\begin{array}{cccc}
\underset{\alpha }{\left( 3,0\right) }, & \underset{\beta }{\left(
2,1\right) }, & \underset{\gamma }{\left( 1,2\right) }, & \underset{\delta }{%
\left( 0,3\right) }%
\end{array}%
.
\end{equation}%
The\textit{\ }$\mathbf{\lambda }$-triality $\mathbb{\tilde{T}}_{\mathbf{%
\lambda }}$ (\ref{1})-(\ref{1'}) of $\mathfrak{so}_{24}$ allows to move
among such four subgroups (within the same group). The theories belonging to
each of such four subgroups will share the same split of the massless
degrees of freedom into bosonic and fermionic ones, respectively specified,
as above, by the numbers $B$ and $F$. Each of such four subgroups is a set
of a varying number of theories, which will be labelled in lowercase Latin
letters : $i$, $ii$, $iii$, etc. Such theories will be connected by the
action of the bosonic map $\mathcal{B}$ (\ref{3})-(\ref{3''}) of $\mathfrak{%
so}_{24}$, and thus they will differ for the content of 5-form $\wedge ^{5}$
and 4-form $\wedge ^{4}$ (potential) fields.

\textit{Modulo} all possibilities arising from the combinations of (\ref%
{poss1})-(\ref{poss3}), the classification of \textit{Monstrous gravity}
theories in $25+1$ space-time dimensions is as follows.

\begin{description}
\item[0] Group $0$ ($\psi $-less theories):%
\begin{eqnarray}
\underset{\text{\textbf{purely bosonic}}}{\underset{\left( B|F\right)
=\left( 196,884|0\right) ,~\lambda \text{-less}}{\alpha }} &:&%
\begin{bmatrix}
& \mathbf{s}_{1} & \mathbf{s}_{2} & \text{\textbf{features}} \\
i & \left( 1,0,3,0,1\right) & \left( 0,16,12,8\right) & \text{{\footnotesize %
bosonic, }}{\footnotesize \wedge }^{{\footnotesize 5}}\text{{\footnotesize %
-less}} \\
ii & ^{\prime \prime } & \left( 1,12,12,8\right) & \text{{\footnotesize %
bosonic}} \\
iii & ^{\prime \prime } & \left( 2,8,12,8\right) & \text{{\footnotesize %
bosonic}} \\
iv & ^{\prime \prime } & \left( 3,4,12,8\right) & \text{{\footnotesize %
bosonic}} \\
v & ^{\prime \prime } & \left( 4,0,12,8\right) & \text{{\footnotesize %
bosonic, }}{\footnotesize \wedge }^{{\footnotesize 4}}\text{{\footnotesize %
-less}}%
\end{bmatrix}
\label{prima} \\
\underset{\left( B|F\right) =\left( 194,836|2,048\right) }{\beta } &:&%
\begin{bmatrix}
& \mathbf{s}_{1} & \mathbf{s}_{2} & \text{\textbf{features}} \\
i & \left( 1,0,2,1,1\right) & \left( 0,16,11,8\right) & {\footnotesize %
\wedge }^{{\footnotesize 5}}\text{{\footnotesize -less}} \\
ii & ^{\prime \prime } & \left( 1,12,11,8\right) & - \\
iii & ^{\prime \prime } & \left( 2,8,11,8\right) & - \\
iv & ^{\prime \prime } & \left( 3,4,11,8\right) & - \\
v & ^{\prime \prime } & \left( 4,0,11,8\right) & {\footnotesize \wedge }^{%
{\footnotesize 4}}\text{{\footnotesize -less}}%
\end{bmatrix}
\\
\underset{\left( B|F\right) =\left( 192,788|4,096\right) }{\gamma } &:&\left[
\begin{array}{cccc}
& \mathbf{s}_{1} & \mathbf{s}_{2} & \text{\textbf{features}} \\
i & \left( 1,0,1,2,1\right) & \left( 0,16,10,8\right) & {\footnotesize %
\wedge }^{{\footnotesize 5}}\text{{\footnotesize -less}} \\
ii & ^{\prime \prime } & \left( 1,12,10,8\right) & - \\
iii & ^{\prime \prime } & \left( 2,8,10,8\right) & - \\
iv & ^{\prime \prime } & \left( 3,4,10,8\right) & - \\
v & ^{\prime \prime } & \left( 4,0,10,8\right) & {\footnotesize \wedge }^{%
{\footnotesize 4}}\text{{\footnotesize -less}}%
\end{array}%
\right] \\
\underset{\left( B|F\right) =\left( 190,740|6,144\right) ,~\wedge ^{1}\text{%
-less}}{\delta } &:&\left[
\begin{array}{cccc}
& \mathbf{s}_{1} & \mathbf{s}_{2} & \text{\textbf{features}} \\
i & \left( 1,0,0,3,1\right) & \left( 0,16,9,8\right) & {\footnotesize \wedge
}^{{\footnotesize 5}}\text{{\footnotesize -less}} \\
ii & ^{\prime \prime } & \left( 1,12,9,8\right) & - \\
iii & ^{\prime \prime } & \left( 2,8,9,8\right) & - \\
iv & ^{\prime \prime } & \left( 3,4,9,8\right) & - \\
v & ^{\prime \prime } & \left( 4,0,9,8\right) & {\footnotesize \wedge }^{%
{\footnotesize 4}}\text{{\footnotesize -less}}%
\end{array}%
\right]
\end{eqnarray}

\item[1] Group $1$ ($\#\psi =1$ theories):%
\begin{eqnarray}
\underset{\left( B|F\right) =\left( 149,780|47,104\right) ,~\lambda \text{%
-less}}{\alpha } &:&\left[
\begin{array}{cccc}
& \mathbf{s}_{1} & \mathbf{s}_{2} & \text{\textbf{features}} \\
i & \left( 1,1,3,0,1\right) & \left( 0,12,10,6\right) & {\footnotesize %
\wedge }^{{\footnotesize 5}}\text{{\footnotesize -less}} \\
ii & ^{\prime \prime } & \left( 1,8,10,6\right) & {\footnotesize -} \\
iii & ^{\prime \prime } & \left( 2,4,10,6\right) & {\footnotesize -} \\
iv & ^{\prime \prime } & \left( 3,0,10,6\right) & {\footnotesize \wedge }^{%
{\footnotesize 4}}\text{{\footnotesize -less}}%
\end{array}%
\right] \\
\underset{\left( B|F\right) =\left( 147,732|49,152\right) }{\beta } &:&\left[
\begin{array}{cccc}
& \mathbf{s}_{1} & \mathbf{s}_{2} & \text{\textbf{features}} \\
i & \left( 1,1,2,1,1\right) & \left( 0,12,9,6\right) & {\footnotesize \wedge
}^{{\footnotesize 5}}\text{{\footnotesize -less}} \\
ii & ^{\prime \prime } & \left( 1,8,9,6\right) & - \\
iii & ^{\prime \prime } & \left( 2,4,9,6\right) & - \\
iv & ^{\prime \prime } & \left( 3,0,9,6\right) & {\footnotesize \wedge }^{%
{\footnotesize 4}}\text{{\footnotesize -less}}%
\end{array}%
\right] \\
\underset{\left( B|F\right) =\left( 145,684|51,200\right) }{\gamma } &:&%
\left[
\begin{array}{cccc}
& \mathbf{s}_{1} & \mathbf{s}_{2} & \text{\textbf{features}} \\
i & \left( 1,1,1,2,1\right) & \left( 0,12,8,6\right) & {\footnotesize \wedge
}^{{\footnotesize 5}}\text{{\footnotesize -less}} \\
ii & ^{\prime \prime } & \left( 1,8,8,6\right) & - \\
iii & ^{\prime \prime } & \left( 2,4,8,6\right) & - \\
iv & ^{\prime \prime } & \left( 3,0,8,6\right) & {\footnotesize \wedge }^{%
{\footnotesize 4}}\text{{\footnotesize -less}}%
\end{array}%
\right] \\
\underset{\left( B|F\right) =\left( 143,636|53,248\right) ,~\wedge ^{1}\text{%
-less}}{\delta } &:&\left[
\begin{array}{cccc}
& \mathbf{s}_{1} & \mathbf{s}_{2} & \text{\textbf{features}} \\
i & \left( 1,1,0,3,1\right) & \left( 0,12,7,6\right) & {\footnotesize \wedge
}^{{\footnotesize 5}}\text{{\footnotesize -less}} \\
ii & ^{\prime \prime } & \left( 1,8,7,6\right) & - \\
iii & ^{\prime \prime } & \left( 2,4,7,6\right) & - \\
iv & ^{\prime \prime } & \left( 3,0,7,6\right) & {\footnotesize \wedge }^{%
{\footnotesize 4}}\text{{\footnotesize -less}}%
\end{array}%
\right]
\end{eqnarray}

\item[2] Group $2$ ($\#\psi =2$ theories):%
\begin{eqnarray}
\underset{\left( B|F\right) =\left( 102,676|94,208\right) ,~\lambda \text{%
-less}}{\alpha } &:&\left[
\begin{array}{cccc}
& \mathbf{s}_{1} & \mathbf{s}_{2} & \text{\textbf{features}} \\
i & \left( 1,2,3,0,1\right) & \left( 0,8,8,4\right) & {\footnotesize \wedge }%
^{{\footnotesize 5}}\text{{\footnotesize -less}} \\
ii & ^{\prime \prime } & \left( 1,4,8,4\right) & {\footnotesize -} \\
iii & ^{\prime \prime } & \left( 2,0,8,4\right) & {\footnotesize \wedge }^{%
{\footnotesize 4}}\text{{\footnotesize -less}}%
\end{array}%
\right] \\
\underset{\left( B|F\right) =\left( 100,628|96,256\right) }{\beta } &:&\left[
\begin{array}{cccc}
& \mathbf{s}_{1} & \mathbf{s}_{2} & \text{\textbf{features}} \\
i & \left( 1,2,2,1,1\right) & \left( 0,8,7,4\right) & {\footnotesize \wedge }%
^{{\footnotesize 5}}\text{{\footnotesize -less}} \\
ii & ^{\prime \prime } & \left( 1,4,7,4\right) & - \\
iii & ^{\prime \prime } & \left( 2,0,7,4\right) & {\footnotesize \wedge }^{%
{\footnotesize 4}}\text{{\footnotesize -less}}%
\end{array}%
\right] \\
\underset{\left( B|F\right) =\left( 98,580|98,304\right) }{\gamma } &:&\left[
\begin{array}{cccc}
& \mathbf{s}_{1} & \mathbf{s}_{2} & \text{\textbf{features}} \\
i & \left( 1,2,1,2,1\right) & \left( 0,8,6,4\right) & {\footnotesize \wedge }%
^{{\footnotesize 5}}\text{{\footnotesize -less}} \\
ii & ^{\prime \prime } & \left( 1,4,6,4\right) & - \\
iii & ^{\prime \prime } & \left( 2,0,6,4\right) & {\footnotesize \wedge }^{%
{\footnotesize 4}}\text{{\footnotesize -less}}%
\end{array}%
\right]  \label{daje} \\
\underset{\left( B|F\right) =\left( 96,532|100,352\right) ,~\wedge ^{1}\text{%
-less}}{\delta } &:&\left[
\begin{array}{cccc}
& \mathbf{s}_{1} & \mathbf{s}_{2} & \text{\textbf{features}} \\
i & \left( 1,2,0,3,1\right) & \left( 0,8,5,4\right) & {\footnotesize \wedge }%
^{{\footnotesize 5}}\text{{\footnotesize -less}} \\
ii & ^{\prime \prime } & \left( 1,4,5,4\right) & - \\
iii & ^{\prime \prime } & \left( 2,0,5,4\right) & {\footnotesize \wedge }^{%
{\footnotesize 4}}\text{{\footnotesize -less}}%
\end{array}%
\right]
\end{eqnarray}

\item[3] Group $3$ ($\#\psi =3$ theories):%
\begin{eqnarray}
\underset{\left( B|F\right) =\left( 55,572|141,312\right) ,~\lambda \text{%
-less}}{\alpha } &:&\left[
\begin{array}{cccc}
& \mathbf{s}_{1} & \mathbf{s}_{2} & \text{\textbf{features}} \\
i & \left( 1,3,3,0,1\right) & \left( 0,4,6,2\right) & {\footnotesize \wedge }%
^{{\footnotesize 5}}\text{{\footnotesize -less}} \\
ii & ^{\prime \prime } & \left( 1,0,6,2\right) & {\footnotesize \wedge }^{%
{\footnotesize 4}}\text{{\footnotesize -less}}%
\end{array}%
\right] \\
\underset{\left( B|F\right) =\left( 53,524|143,360\right) }{\beta } &:&\left[
\begin{array}{cccc}
& \mathbf{s}_{1} & \mathbf{s}_{2} & \text{\textbf{features}} \\
i & \left( 1,3,2,1,1\right) & \left( 0,4,5,2\right) & {\footnotesize \wedge }%
^{{\footnotesize 5}}\text{{\footnotesize -less}} \\
ii & ^{\prime \prime } & \left( 1,0,5,2\right) & {\footnotesize \wedge }^{%
{\footnotesize 4}}\text{{\footnotesize -less}}%
\end{array}%
\right] \\
\underset{\left( B|F\right) =\left( 51,476|145,408\right) }{\gamma } &:&%
\left[
\begin{array}{cccc}
& \mathbf{s}_{1} & \mathbf{s}_{2} & \text{\textbf{features}} \\
i & \left( 1,3,1,2,1\right) & \left( 0,4,4,2\right) & {\footnotesize \wedge }%
^{{\footnotesize 5}}\text{{\footnotesize -less}} \\
ii & ^{\prime \prime } & \left( 1,0,4,2\right) & {\footnotesize \wedge }^{%
{\footnotesize 4}}\text{{\footnotesize -less}}%
\end{array}%
\right] \\
\underset{\left( B|F\right) =\left( 49,428|147,456\right) ,~\wedge ^{1}\text{%
-less}}{\delta } &:&\left[
\begin{array}{cccc}
& \mathbf{s}_{1} & \mathbf{s}_{2} & \text{\textbf{features}} \\
i & \left( 1,3,0,3,1\right) & \left( 0,4,3,2\right) & {\footnotesize \wedge }%
^{{\footnotesize 5}}\text{{\footnotesize -less}} \\
ii & ^{\prime \prime } & \left( 1,0,3,2\right) & {\footnotesize \wedge }^{%
{\footnotesize 4}}\text{{\footnotesize -less}}%
\end{array}%
\right]
\end{eqnarray}

\item[4] Group $4$ ($\#\psi =4$ theories):%
\begin{eqnarray}
\underset{\left( B|F\right) =\left( 8,468|188,416\right) ,~\lambda \text{%
-less}}{\alpha } &:&\left[
\begin{array}{ccc}
\mathbf{s}_{1} & \mathbf{s}_{2} & \text{\textbf{features}} \\
\left( 1,4,3,0,1\right) & \left( 0,0,4,0\right) & {\footnotesize \wedge }^{%
{\footnotesize 5}},{\footnotesize \wedge }^{{\footnotesize 4}},%
{\footnotesize \wedge }^{{\footnotesize 2}}\text{{\footnotesize -less}}%
\end{array}%
\right]  \label{penultima} \\
\underset{\left( B|F\right) =\left( 6,420|190,464\right) }{\beta } &:&\left[
\begin{array}{ccc}
\mathbf{s}_{1} & \mathbf{s}_{2} & \text{\textbf{features}} \\
\left( 1,4,2,1,1\right) & \left( 0,0,3,0\right) & {\footnotesize \wedge }^{%
{\footnotesize 5}},{\footnotesize \wedge }^{{\footnotesize 4}},%
{\footnotesize \wedge }^{{\footnotesize 2}}\text{{\footnotesize -less}}%
\end{array}%
\right] \\
\underset{\left( B|F\right) =\left( 4,372|192,512\right) }{\gamma } &:&\left[
\begin{array}{ccc}
\mathbf{s}_{1} & \mathbf{s}_{2} & \text{\textbf{features}} \\
\left( 1,4,1,2,1\right) & \left( 0,0,2,0\right) & {\footnotesize \wedge }^{%
{\footnotesize 5}},{\footnotesize \wedge }^{{\footnotesize 4}},%
{\footnotesize \wedge }^{{\footnotesize 2}}\text{{\footnotesize -less}}%
\end{array}%
\right] \\
\underset{\left( B|F\right) =\left( 2,324|194,560\right) }{\delta } &:&\left[
\begin{array}{ccc}
\mathbf{s}_{1} & \mathbf{s}_{2} & \text{\textbf{features}} \\
\left( 1,4,0,3,1\right) & \left( 0,0,1,0\right) & {\footnotesize \wedge }^{%
{\footnotesize 5}},{\footnotesize \wedge }^{{\footnotesize 4}},%
{\footnotesize \wedge }^{{\footnotesize 2}},{\footnotesize \wedge }^{%
{\footnotesize 1}}\text{{\footnotesize -less}}%
\end{array}%
\right]  \label{ultima}
\end{eqnarray}
\end{description}

The above classification contains $60$ Monstrous gravity theories, from the
purely bosonic, $\wedge ^{5}$-less, $0.\alpha .i$ theory (\ref{prima}) to
the theory with the highest $F$, i.e. the $4.\delta $ theory (\ref{ultima}).
Note that, since we have imposed $\#g=\#\phi =1$, no purely fermionic
Monstrous gravity can exist. Moreover, as far as linear realizations of
(local) supersymmetry are concerned, Monstrous gravity theories are \textit{%
not} supersymmetric, as it is evident from $B\neq F$ in all cases. It is
also worth remarking that all such theories (but the ones of the group
\textbf{4} (\ref{penultima})-(\ref{ultima})) contain bosonic string theory,
whose (massless, closed string) field content is $\#g=\#\phi =\#\wedge
^{2}=1 $ (see e.g. \cite{bMtheory}), as a subsector.

\section{\label{MMtheory}Monstrous M-theory in $26+1$}

At this point, the natural question arises whether Monstrous gravities
classified above can be uplifted\footnote{%
The possibility of an uplift/oxidation to $26+1$ is far from being trivial,
and when possible, it uniquely fixes the content of the higher dimensional
(massless) spectrum.} to $26+1$ space-time dimensions, in which the massless
little group is $SO_{25}$.

\textit{At least} in one case, namely for the purely bosonic Monstrous
gravity labelled by $0.\alpha .iii$ , the answer to this question is
positive. The field content of such a theory is specified by the following $%
\mathbf{s}_{1}$ and $\mathbf{s}_{2}$, as from (\ref{prima}) :%
\begin{equation}
\underset{\left( B|F\right) =\left( 196,884|0\right) }{0.\alpha .iii}%
:\left\{
\begin{array}{l}
\mathbf{s}_{1}=\left( 1,0,3,0,1\right) ; \\
\\
\mathbf{s}_{2}=\left( 2,8,12,8\right) ,%
\end{array}%
\right.  \label{pre-fieldc}
\end{equation}%
or equivalently (absent fields are not reported):%
\begin{equation}
\begin{array}{ccc}
\text{field} & \mathbf{R}\text{~{\footnotesize of~}}\mathfrak{so}_{24} &
{\footnotesize \#} \\
g: & \mathbf{299} & 1 \\
\wedge ^{1} & \mathbf{24} & 3 \\
\phi : & \mathbf{1} & 1 \\
\wedge ^{5}: & \mathbf{42,504} & 2 \\
\wedge ^{4}: & \mathbf{10,626} & 8 \\
\wedge ^{3}: & \mathbf{2,024} & 12 \\
\wedge ^{2}: & \mathbf{276} & 8%
\end{array}
\label{fieldc}
\end{equation}%
One can indeed realize that all such bosonic massless ($SO_{24}$-covariant)
fields in $25+1$ can be obtained by a KK reduction of the following set of ($%
SO_{25}$-covariant) bosonic massless fields fields in $26+1$ :%
\begin{equation}
\begin{array}{ccccc}
\mathfrak{so}_{25} &  & \mathfrak{so}_{24} & \text{{\footnotesize fields}} &
\\
\underset{\mathbf{324}}{g} & \longrightarrow & \underset{\mathbf{299}}{g}, &
\underset{\mathbf{24}}{\wedge ^{1}}, & \underset{\mathbf{1}}{\phi } \\
\underset{\mathbf{53,130}}{\wedge ^{5}} & \longrightarrow & \underset{%
\mathbf{42,504}}{\wedge ^{5}}, & \underset{\mathbf{10,626}}{\wedge ^{4}} &
\\
\underset{\mathbf{12,650}}{\wedge ^{4}} & \longrightarrow & \underset{%
\mathbf{10,626}}{\wedge ^{4}}, & \underset{\mathbf{2,024}}{\wedge ^{3}} &
\\
\underset{\mathbf{2,300}}{\wedge ^{3}} & \longrightarrow & \underset{\mathbf{%
2,024}}{\wedge ^{3}}, & \underset{\mathbf{276}}{\wedge ^{2}} &  \\
\underset{\mathbf{300}}{\wedge ^{2}} & \longrightarrow & \underset{\mathbf{%
276}}{\wedge ^{2}}, & \underset{\mathbf{24}}{\wedge ^{1}} &
\end{array}
\label{redd}
\end{equation}%
In other words, the (massless) field content (\ref{fieldc}) of the Monstrous
gravity $0.\alpha .iii$ in $25+1$ can be obtained by the $S^{1}$ reduction
of the following (massless) field content in $26+1$:%
\begin{equation}
\begin{array}{ccc}
\text{field} & \mathbf{R}\text{~{\footnotesize of~}}\mathfrak{so}_{25} &
{\footnotesize \#} \\
g: & \mathbf{324} & 1 \\
\wedge ^{5}: & \mathbf{53,130} & 2 \\
\wedge ^{4}: & \mathbf{12,650} & 6 \\
\wedge ^{3}: & \mathbf{2,300} & 6 \\
\wedge ^{2}: & \mathbf{300} & 2%
\end{array}
\label{fieldc'}
\end{equation}

Therefore, we have picked an Einstein gravity theory coupled to $p$-forms,
with $p=2,3,4,5$, in $26+1$ space-time dimensions (that can be coupled to a $%
\mathbf{98,304}$ Rarita-Schwinger field), whose massless spectrum contains $%
196,884$ degrees of freedom that may be acted upon by the Monster group $%
\mathbb{M}$, \textit{at least} after reduction to $D=25+1$, and after
suitable assignment. The assignment is as follows in $D=25+1$: $98,280=(%
\mathbf{42,504}+4\cdot \mathbf{10,626}+6\cdot \mathbf{2,024})+4\cdot \mathbf{%
276}+\mathbf{24}$ to the norm four (i.e., minimal) Leech vectors modulo $%
\mathbb{Z}_{2}$, and hence $196,884=1+299+98,280+98,304$ which corresponds
to the Griess algebra, namely to the sum of the two smallest representations
of $\mathbb{M}$, namely the trivial (singlet) $\mathbf{1}$ and the smallest
non-trivial one $\mathbf{196,883}$. Such a theory will be henceforth named
\textit{Monstrous M-theory}, or simply \textit{M}$^{2}$\textit{-theory}.
Note that the disentangling of the $196,884$ degrees of freedom into $%
\mathbf{196,883}\oplus \mathbf{1}$ occurs only when reducing the theory to $%
25+1$, in which case the dilaton $\phi $ is identified with the singlet of $%
\mathbb{M}$ : in other words, \textit{the (observation which firstly hinted
the) Monstrous Moonshine} \cite{conway79}\textit{\ is crucially related to
the }$S^{1}$\textit{\ compactification of M}$^{2}$\textit{-theory down to }$%
25+1$\textit{\ space-time dimensions}.

\subsection{\label{Lagr}Lagrangian(s) for Bosonic Monstrous M-theory}

\textit{A priori}, the purely bosonic $196,884$-dimensional degrees of
freedom of the massless spectrum of M$^{2}$-theory can be realized in
various ways at the Lagrangian level. Here, within the framework defined
above, we will attempt at writing down a general Lagrangian for bosonic part
of M$^{2}$-theory.

We start and label the massless fields of M$^{2}$-theory, given by (\ref%
{fieldc'}), as follows\footnote{%
It is amusing to note that the $p$-form (potentials) content of M$^{2}$%
-theory follows from a pair of 5-form (potentials) of $SO_{28}$, which is
the massless little group in $30$ dimensions. Thus, the bosonic
non-gravitational content of M$^{2}$-theory descends from a pair of massless
4-branes in $D=s+t=30$,or better from a self-dual pair of massless $p$-form
potentials in $D=30$, namely from a $5$-form and its dual $23$-form
potentials, respectively related to \textit{massless} 4-brane and its dual
22-brane in $D=30$.} :%
\begin{equation}
\begin{array}{ccc}
\text{field} & \text{label} & {\footnotesize \#} \\
g: & g_{\mu \nu } & 1 \\
\wedge ^{5}: & C_{\lambda \mu \nu \rho \sigma }^{(5)A} & 2 \\
\wedge ^{4}: & C_{\lambda \mu \nu \rho }^{(4)i} & 6 \\
\wedge ^{3}: & C_{\lambda \mu \nu }^{(3)i} & 6 \\
\wedge ^{2}: & C_{\lambda \mu }^{(2)A} & 2%
\end{array}%
\end{equation}%
The uppercase Latin indices take values $1,2$, whereas the lowercase Latin
indices run $1,2,..,6$. A general definition of the field strengths reads%
\begin{equation}
\begin{array}{l}
G^{(3)A}:=dC^{(2)A}+\mathbf{A}_{j}^{A}C^{(3)j}; \\
G^{(4)i}:=dC^{(3)i}+\mathbf{B}_{(AB)}^{i}C^{(2)A}\wedge C^{(2)B}+\mathbf{C}%
_{ij}C^{(4)j}; \\
G^{(5)i}:=dC^{(4)i}+\mathbf{D}_{Aj}^{i}C^{(2)A}\wedge C^{(3)j}+\mathbf{E}%
_{A}^{i}C^{(5)A}; \\
G^{(6)A}:=dC^{(5)A}+\mathbf{F}_{\left( BCD\right) }^{A}C^{(2)B}\wedge
C^{(2)C}\wedge C^{(2)D}+\mathbf{G}_{ij}^{A}C^{(3)i}\wedge C^{(3)j}+\mathbf{H}%
_{Bi}^{A}C^{(4)i}\wedge C^{(2)B},%
\end{array}
\label{fs}
\end{equation}%
where the uppercase bold Latin tensors are constant\footnote{%
All (uppercase and calligraphic) Latin tensors introduced in (\ref{L})-(\ref%
{L''}) are constant, because there is no scalar field in the (massless)
spectrum of the theory.}, and they are possibly given by suitable
representation theoretic projectors\footnote{%
Here we will not analyze possible characterizations of such tensor as
(invariant) projectors. We confine ourselves to remark that, in a very
simple choice of covariance (namely, $A=1,2$ and $i=1,2,...,6$ running over
the spin-$1/2$ and spin-$5/2$ representations $\mathbf{2}$ and $\mathbf{6}$
of $\mathfrak{sl}_{2}$), most of them vanish.}.

Then, a general Lagrangian density can be written as%
\begin{eqnarray}
\mathcal{L} &=&R-\frac{1}{2\cdot 3!}\mathcal{A}_{AB}G^{(3)A}\cdot G^{(3)B}-%
\frac{1}{2\cdot 4!}\mathcal{B}_{ij}G^{(4)i}\cdot G^{(4)j}  \notag \\
&&-\frac{1}{2\cdot 5!}\mathcal{C}_{ij}G^{(5)i}\cdot G^{(5)j}-\frac{1}{2\cdot
6!}\mathcal{D}_{AB}G^{(6)A}\cdot G^{(6)B}+\mathcal{L}_{CS\text{-like}},
\label{L}
\end{eqnarray}%
where the calligraphic Latin constant tensors are (symmetric and) positive
definite in order for all kinetic terms of $p$-forms to be consistent. A
minimal, Maxwell-like choice is $\mathcal{A}_{AB}=\mathcal{D}_{AB}=\delta
_{AB}$ and $\mathcal{B}_{ij}=\mathcal{C}_{ij}=\delta _{ij}$, such that (\ref%
{L}) simplifies down to
\begin{eqnarray}
\mathcal{L} &=&R-\frac{1}{2\cdot 3!}\sum_{A=1}^{2}G_{\mu \nu \rho
}^{(3)A}G^{(3)A|\mu \nu \rho }-\frac{1}{2\cdot 4!}\sum_{i=1}^{6}G_{\lambda
\mu \nu \rho }^{(4)i}G^{(4)i|\lambda \mu \nu \rho }  \notag \\
&&-\frac{1}{2\cdot 5!}\sum_{i=1}^{6}G_{\lambda \mu \nu \rho \sigma
}^{(5)i}G^{(5)i|\lambda \mu \nu \rho \sigma }-\frac{1}{2\cdot 6!}%
\sum_{A=1}^{2}G_{\lambda \mu \nu \rho \sigma \tau }^{(6)A}G^{(6)A|\lambda
\mu \nu \rho \sigma \tau }+\mathcal{L}_{CS\text{-like}}.  \label{L'}
\end{eqnarray}%
The \textquotedblleft topological\textquotedblright , \textquotedblleft
Chern-Simons-like\textquotedblright\ Lagrangian occurring in (\ref{L}) and (%
\ref{L'}) is composed by a number of \textit{a priori} non-vanishing terms,
such as for instance :%
\begin{eqnarray}
\sqrt{|g|}\mathcal{L}_{CS} &=&\epsilon \mathcal{E}%
_{3}^{ABCDi}G_{A}^{(6)}G_{B}^{(6)}G_{C}^{(6)}G_{D}^{(6)}C_{i}^{(3)}  \notag
\\
&&+\epsilon \mathcal{I}%
_{2}^{ijklmA}G_{i}^{(5)}G_{j}^{(5)}G_{k}^{(5)}G_{l}^{(5)}G_{m}^{(5)}C_{A}^{(2)}+\dots
\notag \\
&&+\epsilon \mathcal{S}%
_{3}^{ijklmnp}G_{i}^{(4)}G_{j}^{(4)}G_{k}^{(4)}G_{l}^{(4)}G_{m}^{(4)}G_{n}^{(4)}C_{p}^{(3)}+\dots
\notag \\
&&+\epsilon \mathcal{W}%
_{3}^{ABCDEFGHi}G_{A}^{(3)}G_{B}^{(3)}G_{C}^{(3)}G_{D}^{(3)j}G_{E}^{(3)}G_{F}^{(3)}G_{G}^{(3)}G_{H}^{(3)}C_{i}^{(3)}+\dots
\label{L''}
\end{eqnarray}%
where $\epsilon $ denotes the Ricci-Levi-Civita tensor in $26+1$, and the
full Lagrangian is shown in Appendix \eqref{monsterCS}. We leave the study
of the constant tensors $\mathbf{A}$, ..., $\mathbf{H}$, $\mathcal{A},...,%
\mathcal{D}$, and $\mathcal{E},...,\mathcal{W}$ resp. in (\ref{fs}), (\ref{L}%
), and (\ref{L''}) (as well as others occurring in App. \eqref{monsterCS})
for further future work.

It is immediate to realize that M$^{2}$-theory includes Horowitz and
Susskind's bosonic M-theory \cite{bMtheory} as a truncation; indeed, by
setting%
\begin{equation}
\begin{array}{l}
C^{(2)A}=0; \\
C^{(3)i}=\delta ^{i1}C; \\
C^{(4)i}=0; \\
C^{(5)A}=0,%
\end{array}%
\end{equation}%
one obtains ($F=dC$)%
\begin{equation}
\mathcal{L}=R-\frac{1}{2\cdot 4!}F^{2},
\end{equation}%
which is the Lagrangian of the bosonic string theory discussed by Susskind
and Horowitz in \cite{bMtheory}.

Finally, we observe that a Scherk-Schwarz reduction of the Lagrangian (\ref%
{L}) to $25+1$ would provide a quite general Lagrangian for the $0.\alpha
.iii$ Monster (dilatonic, Einstein) gravity; we leave this task for further
future work.

\subsection{\label{BBFF}$B=F$ in $26+1$}

Remarkably, a certain subsector of M$^{2}$-theory, when coupled to an $h=3/2$
Rarita-Schwinger field, exhibits $B=F$, which is a necessary condition for
(linearly realized, conventional) supersymmetry to hold. Such a subsector is
given by\footnote{%
Analogously to what observed in Sec. \ref{Lagr}, it is amusing to observe
that the bosonic content of the $B=F$ sector of $26+1$ M$^{2}$-theory (which
we are tempted to conjecture to be $\mathcal{N}=1$, $D=26+1$ supergravity;
see further below) derives from a single $5$-form potential, corresponding
to a \textit{massless} $4$-brane, in $D=30$, complemented by a
\textquotedblleft transmutation\textquotedblright\ of the $2$-form potential
$\mathbf{300}$ of $\mathfrak{so}_{25}$ into the rank-$2$ symmetric traceless
tensor (graviton) $\mathbf{324}$ of $\mathfrak{so}_{25}$, namely by the
replacement of a massless string (1-brane) with a massless graviton in $%
D=26+1$.}%
\begin{equation}
\begin{array}{ccc}
\text{field} & \mathbf{R}\text{~{\footnotesize of~}}\mathfrak{so}_{25} &
{\footnotesize \#} \\
g: & \mathbf{324} & 1 \\
\wedge ^{5}: & \mathbf{53,130} & 1 \\
\wedge ^{4}: & \mathbf{12,650} & 3 \\
\wedge ^{3}: & \mathbf{2,300} & 3 \\
\wedge ^{2}: & \mathbf{300} & 0%
\end{array}
\label{sub}
\end{equation}%
Thus, when coupled to a an $h=3/2$ RS field $\psi $ (fitting the $\mathbf{%
98,304}$ irreducible representation of $\mathfrak{so}_{25}$, with Dynkin
labels $\left( 1,0^{10},1\right) $), the resulting theory has\footnote{%
Again, bosonic M-theory \cite{bMtheory} trivially is a subsector of (the
purely bosonic sector of) such a theory in $26+1$.}%
\begin{equation}
B=F=98,304.  \label{BF}
\end{equation}%
By recalling (\ref{redd}) and observing that the massless RS field branches
from $26+1$ to $25+1$ as%
\begin{equation}
\underset{\mathfrak{so}_{25}~\text{repr.}}{\underbrace{\underset{\mathbf{%
\psi }}{\mathbf{98,304}}}}=\underset{\mathfrak{so}_{24}\text{ reprs.}}{%
\underbrace{\underset{\mathbf{\psi }}{\mathbf{47,104}}\oplus \underset{%
\mathbf{\psi }^{\prime }}{\mathbf{47,104}^{\prime }}\oplus \underset{\mathbf{%
\lambda }}{\mathbf{2,048}}\oplus \underset{\mathbf{\lambda }^{\prime }}{%
\mathbf{2,048}^{\prime }}}},
\end{equation}%
the subsector of M$^{2}$-theory with $B=F=98,304$ gives rise to the
following massless spectrum, when reduced to $25+1$ :%
\begin{equation}
\begin{array}{ccc}
\text{field} & \mathbf{R}\text{~{\footnotesize of~}}\mathfrak{so}_{24} &
{\footnotesize \#} \\
g: & \mathbf{299} & 1 \\
\psi : & \mathbf{47,104} & 2\equiv \left( \psi \oplus \psi ^{\prime }\right)
\\
\wedge ^{1}: & \mathbf{24} & 1 \\
\lambda : & \mathbf{2,048} & 2\equiv \left( \lambda \oplus \lambda ^{\prime
}\right) \\
\varphi : & \mathbf{1} & 1 \\
\wedge ^{5}: & \mathbf{42,504} & 1 \\
\wedge ^{4}: & \mathbf{10,626} & 4 \\
\wedge ^{3}: & \mathbf{2,024} & 6 \\
\wedge ^{2}: & \mathbf{276} & 3%
\end{array}
\label{subb}
\end{equation}%
By recalling the treatment of Sec. \ref{MGra}, one can recognize (\ref{subb}%
) as a subsector (in which (\ref{BF}) holds) of the Monstrous gravity $%
2.\gamma .ii$ in (\ref{daje}), simply obtained by decreasing $\#\wedge ^{2}$
from $4$ to $3$.

Other subsectors of Monstrous gravity theories in $25+1$ exist such that $%
B=F $; below, we list some of them :%
\begin{equation}
\begin{array}{cccccc}
&  &  & \mathbf{s}_{1} & \mathbf{s}_{2} & B=F \\
0. & \gamma .i-iv &  & \left( 0,0,1,1,0\right) & \left( 0,0,1,0\right) &
2,048 \\
&  &  &  &  &  \\
2. & \left\{
\begin{array}{l}
\alpha .i-ii \\
\beta .i-ii \\
\gamma .i-ii \\
\delta .i-ii%
\end{array}%
\right. &  & \left( 0,2,0,0,0\right) & \left( 1,4,4,4\right) & 94,208 \\
&  &  &  &  &  \\
2. & \left\{
\begin{array}{c}
\gamma .ii \\
\delta .ii%
\end{array}%
\right. &  &
\begin{array}{c}
\left( 0,2,1,1,0\right) \\
\left( 1,2,0,1,1\right)%
\end{array}
&
\begin{array}{c}
\left( 1,4,5,4\right) \\
\left( 1,4,5,3\right)%
\end{array}
& 96,256%
\end{array}
\label{subbb}
\end{equation}%
Note that, among the $B=F$ subsectors in $25+1$ reported above, only (\ref%
{subb}) and the second in the last line of (\ref{subbb}) (i.e., the
subsector of the $2.\delta .ii$ Monstrous gravity) contain gravity.

\subsubsection{\label{SS}$\mathcal{N}=1$ Supergravity in $26+1$?}

As pointed out, $B=F$ is a necessary but not sufficient condition for
(linearly realized, local, conventional) supersymmetry to hold. It is thus
tantalizing to conjecture that the theory in $26+1$ with massless spectrum (%
\ref{sub}) and one Rarita-Schwinger field $\psi $ is actually an $\mathcal{N}%
=1$ supergravity theory.

Inspired by M-theory\footnote{%
Throughout our treatment, we refer to the conventions used in Sec. 22 of
\cite{Ortin-book}.} (i.e., $\mathcal{N}=1$ supergravity) in $10+1$, and
exploiting a truncation of the purely bosonic Lagrangians discussed in Sec. %
\ref{Lagr} (the capped lowercase Latin indices run $\hat{\imath}=1,2,3$
throughout), one can write down a tentative Lagrangian for the would-be $%
\mathcal{N}=1$ supergravity in $26+1$ :%
\begin{eqnarray}
\mathcal{L} &=&R-\frac{1}{2\cdot 4!}\sum_{\hat{\imath}=1}^{3}G^{(4)\hat{%
\imath}}\cdot G^{(4)\hat{\imath}}-\frac{1}{2\cdot 5!}\sum_{\hat{\imath}%
=1}^{3}G^{(5)\hat{\imath}}\cdot G^{(5)\hat{\imath}}-\frac{1}{2\cdot 6!}%
G^{(6)}\cdot G^{(6)}+\mathcal{L}_{CS\text{-like}}  \notag \\
&&-\mathbf{a}\frac{i}{2}\overline{\psi }_{\mu }\Gamma ^{\mu \nu \rho }\nabla
_{\nu }\left( \frac{\omega +\tilde{\omega}}{2}\right) \psi _{\rho }  \notag
\\
&&+\sum_{\hat{\imath}=1}^{3}\mathbf{b}_{\hat{\imath}}\overline{\psi }_{\mu
}\Gamma ^{\lbrack \mu }\Gamma ^{(4)}\Gamma ^{\nu ]}\psi _{\nu }\cdot \left(
G^{(4)\hat{\imath}}+\tilde{G}^{(4)\hat{\imath}}\right)  \notag \\
&&+\sum_{\hat{\imath}=1}^{3}\mathbf{c}_{\hat{\imath}}\overline{\psi }_{\mu
}\Gamma ^{\lbrack \mu }\Gamma ^{(5)}\Gamma ^{\nu ]}\psi _{\nu }\left( G^{(5)%
\hat{\imath}}+\tilde{G}^{(5)\hat{\imath}}\right)  \notag \\
&&+\mathbf{d}\overline{\psi }_{\mu }\Gamma ^{\lbrack \mu }\Gamma
^{(6)}\Gamma ^{\nu ]}\psi _{\nu }\cdot \left( G^{(6)}+\tilde{G}^{(6)}\right)
,  \label{L-sugra}
\end{eqnarray}%
where%
\begin{equation}
\Gamma ^{(4)}\cdot G^{(4)\hat{\imath}}=\Gamma _{\alpha \beta \gamma \delta
}G^{(4)\hat{\imath}|\alpha \beta \gamma \delta }\text{, etc.}  \label{def}
\end{equation}%
and, upon truncation of (\ref{fs}) resp. (\ref{L''}),%
\begin{equation}
\begin{array}{l}
G^{(4)\hat{\imath}}:=dC^{(3)\hat{\imath}}+\mathbf{C}_{\hat{\imath}\hat{\jmath%
}}C^{(4)\hat{\jmath}}; \\
G^{(5)\hat{\imath}}:=dC^{(4)\hat{\imath}}+\mathbf{E}^{\hat{\imath}}C^{(5)};
\\
G^{(6)}:=dC^{(5)}+\mathbf{G}_{\hat{\imath}\hat{\jmath}}C^{(3)\hat{\imath}%
}\wedge C^{(3)\hat{\jmath}};%
\end{array}
\label{fs'}
\end{equation}%
\begin{eqnarray}
\sqrt{|g|}\mathcal{L}_{CS\text{-like}} &=&\epsilon \mathcal{E}_{\hat{\imath}%
}G^{(6)}G^{(6)}G^{(6)}G^{(6)}C^{(3)\hat{\imath}}  \notag \\
&&+\epsilon \mathcal{G}_{\hat{\imath}\hat{\jmath}\hat{k}\hat{l}\hat{m}\hat{n}%
\hat{p}}G^{(4)\hat{\imath}}G^{(4)\hat{\jmath}}G^{(4)\hat{k}}G^{(4)\hat{l}%
}G^{(4)\hat{m}}G^{(4)\hat{n}}C^{(3)\hat{p}}  \notag \\
&&+\epsilon \mathcal{H}_{\hat{\imath}\hat{\jmath}\hat{k}\hat{l}%
}G^{(6)}G^{(6)}G^{(4)\hat{\imath}}G^{(4)\hat{\jmath}}G^{(4)\hat{k}}C^{(3)%
\hat{l}}  \notag \\
&&+\epsilon \mathcal{I}_{\hat{\imath}\hat{\jmath}\hat{k}\hat{l}\hat{m}%
}G^{(6)}G^{(5)\hat{\imath}}G^{(5)\hat{\jmath}}G^{(4)\hat{k}}G^{(4)\hat{l}%
}C^{(3)\hat{m}}  \notag \\
&&+\epsilon \mathcal{J}_{\hat{\imath}}G^{(6)}G^{(6)}G^{(6)}G^{(5)\hat{\imath}%
}C^{(4)\hat{\jmath}}  \notag \\
&&+\epsilon \mathcal{K}_{\hat{\imath}}G^{(6)}G^{(6)}G^{(6)}G^{(4)\hat{\imath}%
}C^{(5)}  \notag \\
&&+\epsilon \mathcal{L}_{\hat{\imath}\hat{\jmath}\hat{k}\hat{l}\hat{m}%
}G^{(6)}G^{(5)\hat{\imath}}G^{(4)\hat{\jmath}}G^{(4)\hat{k}}G^{(4)\hat{l}%
}C^{(4)\hat{m}}  \notag \\
&&+\epsilon \mathcal{M}_{\hat{\imath}\hat{\jmath}}G^{(6)}G^{(6)}G^{(5)\hat{%
\imath}}G^{(5)\hat{\jmath}}C^{(5)}  \notag \\
&&+\epsilon \mathcal{N}_{\hat{\imath}\hat{\jmath}\hat{k}\hat{l}}G^{(6)}G^{(4)%
\hat{\imath}}G^{(4)\hat{\jmath}}G^{(4)\hat{k}}G^{(4)\hat{l}}C^{(5)}  \notag
\\
&&+\epsilon \mathcal{O}_{\hat{\imath}\hat{\jmath}\hat{k}\hat{l}\hat{m}\hat{n}%
}G^{(5)\hat{\imath}}G^{(5)\hat{\jmath}}G^{(5)\hat{k}}G^{(5)\hat{l}}G^{(4)%
\hat{m}}C^{(3)\hat{n}}  \notag \\
&&+\epsilon \mathcal{P}_{\hat{\imath}\hat{\jmath}\hat{k}\hat{l}\hat{m}\hat{n}%
}G^{(5)\hat{\imath}}G^{(5)\hat{\jmath}}G^{(5)\hat{k}}G^{(4)\hat{l}}G^{(4)%
\hat{m}}C^{(4)\hat{n}}.  \label{L'''}
\end{eqnarray}%
Moreover,%
\begin{equation}
\begin{array}{l}
\tilde{G}^{(4)\hat{\imath}}:=G^{(4)\hat{\imath}}+\mathbf{e}^{\hat{\imath}}%
\overline{\psi }\Gamma ^{(2)}\psi ; \\
\tilde{G}^{(5)\hat{\imath}}:=G^{(5)\hat{\imath}}+\mathbf{f}^{\hat{\imath}}%
\overline{\psi }\Gamma ^{(3)}\psi ; \\
\tilde{G}^{(6)}:=G^{(6)}+\mathbf{g}\overline{\psi }\Gamma ^{(4)}\psi%
\end{array}
\label{fs''}
\end{equation}%
are the would-be supercovariant field strengths, and%
\begin{equation}
\nabla _{\mu }\left( \omega \right) \psi _{\nu }:=\partial _{\mu }\psi _{\nu
}+\mathbf{h}\omega _{\mu }^{ab}\Gamma _{ab}\psi _{\nu }
\end{equation}%
is the covariant derivative with%
\begin{eqnarray}
\tilde{\omega}_{\mu }^{ab} &:&=\omega _{\mu }^{ab}+i\mathbf{l}\overline{\psi
}_{\alpha }\Gamma _{\mu }^{ab\alpha \beta }\psi _{\beta }; \\
\omega _{\mu }^{ab} &:&=\omega _{\mu }^{ab}\left( e\right) +K_{\mu }^{ab}; \\
K_{\mu }^{ab} &:&=i\left[ \mathbf{m}\overline{\psi }_{\alpha }\Gamma _{\mu
}^{ab\alpha \beta }\psi _{\beta }+\mathbf{n}\left( \overline{\psi }_{\mu
}\Gamma ^{b}\psi ^{a}-\overline{\psi }_{\mu }\Gamma ^{a}\psi ^{b}+\overline{%
\psi }^{b}\Gamma _{\mu }\psi ^{a}\right) \right] .  \label{deff}
\end{eqnarray}%
We can therefore formulate the following\smallskip

\textbf{Conjecture\smallskip }

The Lagrangian (\ref{L-sugra}) should be invariant under the following local
supersymmetry transformations with parameter $\varepsilon $ (a Majorana
spinor) :%
\begin{eqnarray}
\delta _{\varepsilon }e_{\mu }^{a} &=&-\frac{i}{2}\bar{\varepsilon}\Gamma
^{a}\psi _{\mu };  \label{susy-1} \\
\delta _{\varepsilon }\psi _{\mu } &=&\mathbf{p}\nabla _{\mu }\left( \tilde{%
\omega}\right) \varepsilon +\sum_{\hat{\imath}=1}^{3}\mathbf{q}_{\hat{\imath}%
}\left( \Gamma _{~~~~~\mu }^{\alpha \beta \gamma \delta }+\mathbf{r}_{\hat{%
\imath}}\Gamma ^{\beta \gamma \delta }\delta _{\mu }^{\alpha }\right)
\varepsilon \tilde{G}_{\alpha \beta \gamma \delta }^{(4)\hat{\imath}}  \notag
\\
&&+\sum_{\hat{\imath}=1}^{3}\mathbf{s}_{\hat{\imath}}\left( \Gamma
_{~~~~~~\mu }^{\alpha \beta \gamma \delta \rho }+\mathbf{t}_{\hat{\imath}%
}\Gamma ^{\beta \gamma \delta \rho }\delta _{\mu }^{\alpha }\right)
\varepsilon \tilde{G}_{\alpha \beta \gamma \delta \rho }^{(5)\hat{\imath}}+%
\mathbf{u}\left( \Gamma _{~~~~~~~\mu }^{\alpha \beta \gamma \delta \rho
\sigma }+\mathbf{v}\Gamma ^{\beta \gamma \delta \rho \sigma }\delta _{\mu
}^{\alpha }\right) \varepsilon \tilde{G}_{\alpha \beta \gamma \delta \rho
\sigma }^{(6)}; \\
\delta _{\varepsilon }C_{\mu \nu \rho }^{(3)\hat{\imath}} &=&\mathbf{w}^{%
\hat{\imath}}\bar{\varepsilon}\Gamma _{\lbrack \mu \nu }\psi _{\rho ]}; \\
\delta _{\varepsilon }C_{\mu \nu \rho \sigma }^{(4)\hat{\imath}} &=&\mathbf{x%
}^{\hat{\imath}}\bar{\varepsilon}\Gamma _{\lbrack \mu \nu \rho }\psi
_{\sigma ]}; \\
\delta _{\varepsilon }C_{\mu \nu \rho \sigma \tau }^{(5)} &=&\mathbf{y}\bar{%
\varepsilon}\Gamma _{\lbrack \mu \nu \rho \rho }\psi _{\tau ]}.
\label{susy-5}
\end{eqnarray}

To prove (or disprove) the invariance of the Lagrangian (\ref{L-sugra})
(with definitions (\ref{def})-(\ref{deff})) under the local supersymmetry
transformations (\ref{susy-1})-(\ref{susy-5}), and thus fixing the real
parameters $\mathbf{a}$,...,$\mathbf{y}$ as well as the tensors $\mathbf{C}$%
, $\mathbf{E}$, $\mathbf{G}$ and $\mathcal{E}$, $\mathcal{G}$, seems a
formidable task, which deserves to be pursued in a separate paper.

Under dimensional reduction to $25+1$, one would then get a would-be type
IIA $\mathcal{N}=(1,1)$ supergravity theory, with massless spectrum (\ref%
{subb}); as observed above, this would correspond to a suitable truncation
of the Monstrous gravity $2.\gamma .ii$ in (\ref{daje}), in which $\#\wedge
^{2}$ decreases from $4$ to $3$; again, we leave the investigation of
interesting task for further future work. 

\section{\label{Coho}Cohomological construction of lattices : from $%
\mathfrak{e}_{8}$ to the Leech lattice}

Let us consider the following (commutative) diagram, starting from the Lie
algebra $\mathfrak{e}_{8}$,

\begin{equation}
\begin{array}{ccccc}
&  & \mathfrak{a}_{8}\oplus \wedge ^{3}\oplus \overline{\wedge ^{3}} &  &
\\
& \nearrow &  & \searrow &  \\
\mathfrak{e}_{8} &  &  &  & \mathfrak{b}_{4}\oplus g\oplus \wedge ^{3}\oplus
\ast \wedge ^{3} \\
& \searrow &  & \nearrow &  \\
&  & \mathfrak{d}_{8}\oplus \mathbf{\lambda } &  &
\end{array}
\label{jazz}
\end{equation}%
where $g\equiv S_{0}^{2}$, as above, denotes the $D=10+1$ graviton
representation (which has been related to \textquotedblleft
super-Ehlers\textquotedblright\ embeddings in \cite{superehlers}), and $\ast
$ stands for the Hodge dual ($\ast \wedge ^{p}:=\wedge ^{D-p}$). Thus, the
number $\#\mathfrak{e}_{8}$ of roots of the $\mathfrak{e}_{8}$ root lattice
reads
\begin{equation}
\underset{240}{\#\mathfrak{e}_{8}}=\underset{248}{\dim \mathfrak{e}_{8}}-8=(%
\underset{32}{\dim \mathfrak{b}_{4}-4)}+(\underset{40}{\dim \mathbf{g}-4}%
)+\dim (\underset{84\cdot 2}{\wedge ^{3}\oplus \ast \wedge ^{3}}).
\end{equation}%
Therefore, the number $\#\mathfrak{e}_{8}^{+}$ of positive roots of $%
\mathfrak{e}_{8}$ is%
\begin{equation}
\#\mathfrak{e}_{8}^{+}=\frac{1}{2}\left( (\underset{32}{\dim \mathfrak{b}%
_{4}-4)}+(\underset{40}{\dim \mathbf{g}-4})\right) +\underset{84}{\dim
\wedge ^{3}}=120.  \label{coho}
\end{equation}%
Note that it also holds that%
\begin{equation}
\frac{1}{2}\left( (\underset{32}{\dim \mathfrak{b}_{4}-4)}+(\underset{40}{%
\dim \mathbf{g}-4})\right) =\dim \mathfrak{b}_{4}.
\end{equation}%
It should be also remarked that $120=\binom{10}{3}$, i.e. it matches the
number of degrees of freedom of a massless 3-form potential in $11+1$
space-dimensions; indeed, a massless 3-form in $11+1$ (corresponding to $%
\wedge ^{3}$ of the little group $SO_{10}$) gives rise to a massless 3-form
and a massless 2-form in $10+1$ (corresponding to $\wedge ^{3}\oplus \wedge
^{2}\simeq \wedge ^{3}\oplus \mathfrak{b}_{4}$ of the little group $SO_{9}$%
).\medskip

The case of $\mathfrak{e}_{8}$ is peculiar, because the closure (as well as
the commutativity) of the diagram (\ref{jazz}) relies on the existence of
the \textquotedblleft anomalous\textquotedblright\ embedding%
\begin{equation}
\begin{array}{c}
\mathfrak{d}_{8}\supset \mathfrak{b}_{4}; \\
\mathbf{16}=\mathbf{16}\equiv \mathbf{\lambda },%
\end{array}
\label{anem}
\end{equation}%
where $\mathbf{\lambda }$ is the spinor representation.

By replacing $\mathfrak{b}_{4}$ and $\wedge ^{3}$ respectively as follows,%
\begin{eqnarray}
\mathfrak{b}_{4} &\rightarrow &\mathfrak{b}_{12}; \\
\wedge ^{3} &\rightarrow &\wedge ^{5}\oplus 3\cdot \wedge ^{4}\oplus 3\cdot
\wedge ^{3},
\end{eqnarray}%
one can define the \textquotedblleft Leech algebra\textquotedblright\ $%
\mathfrak{L}_{24}$ in analogy with $\mathfrak{e}_{8}$ (albeit with $D=26+1$
graviton $g$), through the following diagram :
\begin{equation}
\begin{array}{ccccc}
&  &
\begin{array}{l}
\mathfrak{a}_{24}\oplus \left( \wedge ^{5}\oplus 3\cdot \wedge ^{4}\oplus
3\cdot \wedge ^{3}\right) \\
\oplus \left( \overline{\wedge ^{5}}\oplus 3\cdot \overline{\wedge ^{4}}%
\oplus 3\cdot \overline{\wedge ^{3}}\right)%
\end{array}
&  &  \\
& \nearrow &  & \searrow &  \\
\mathfrak{L}_{24} &  &  &  &
\begin{array}{l}
\mathfrak{b}_{12}\oplus g\oplus \left( \wedge ^{5}\oplus 3\cdot \wedge
^{4}\oplus 3\cdot \wedge ^{3}\right) \\
\oplus \left( \ast \left( \wedge ^{5}\oplus 3\cdot \wedge ^{4}\oplus 3\cdot
\wedge ^{3}\right) \right)%
\end{array}
\\
& \searrow &  & \nearrow &  \\
&  & ? &  &
\end{array}
\label{jazz-2}
\end{equation}%
The question mark in (\ref{jazz-2}) occurs because \textit{there is no
analogue of the \textquotedblleft anomalous\textquotedblright\ embedding (%
\ref{anem}) for }$\mathfrak{L}_{24}$. Thus, it holds that%
\begin{eqnarray}
\underset{196,560}{\#\mathfrak{L}_{24}} &=&\underset{196,584}{\dim \mathfrak{%
L}_{24}}-24  \notag \\
&=&(\underset{288}{\dim \mathfrak{b}_{12}-12})+(\underset{312}{\dim \mathbf{g%
}-12})+\dim (\underset{2\cdot \left( 53,130+3\cdot 12,650+3\cdot
2,300\right) }{\wedge ^{5}\oplus 3\cdot \wedge ^{4}\oplus 3\cdot \wedge
^{3}+\ast \left( \wedge ^{5}\oplus 3\cdot \wedge ^{4}\oplus 3\cdot \wedge
^{3}\right) })  \notag \\
&=&196,560,
\end{eqnarray}%
where $\#\mathfrak{L}_{24}$ denotes the number of minimal, non-trivial
vectors (of norm $4$) of the Leech lattice $\Lambda _{24}$. Therefore, the $%
\mathbb{Z}_{2}$-modded number of minimal, non-trivial vectors of $\Lambda
_{24}$ is%
\begin{equation}
\#\mathfrak{L}_{24}^{+}=\frac{1}{2}\left( (\underset{288}{\dim \mathfrak{b}%
_{12}-12})+(\underset{312}{\dim \mathbf{g}-12})\right) +\dim (\underset{%
53,130+3\cdot 12,650+3\cdot 2,300}{\wedge ^{5}\oplus 3\cdot \wedge
^{4}\oplus 3\cdot \wedge ^{3}})=98,280,  \label{coho-2}
\end{equation}%
which is the number entering the construction of the smallest non-trivial
representation of the Monster group $\mathbb{M}$ (cfr. \cite{Conway-square}%
). Note that it also holds that%
\begin{equation}
\frac{1}{2}\left( (\underset{288}{\dim \mathfrak{b}_{12}-12})+(\underset{312}%
{\dim \mathbf{g}-12})\right) =\dim \mathfrak{b}_{12}.
\end{equation}%
It should moreover be also remarked that $98,280=\binom{28}{5}$, i.e. it
matches the number of degrees of freedom of a massless 5-form potential in $%
D=29+1$ space-dimensions; indeed, it can be checked that a massless 5-form
potential in $D=29+1$ (corresponding to $\wedge ^{5}$ of the little group $%
SO_{28}$) gives rise to 1 massless 5-form, 3 massless 4-forms, 3 massless
3-forms and 1 massless 2-form in $26+1$ (corresponding to $\left( \wedge
^{5}\oplus 3\cdot \wedge ^{4}\oplus 3\cdot \wedge ^{3}\right) \oplus \wedge
^{2}\simeq \left( \wedge ^{5}\oplus 3\cdot \wedge ^{4}\oplus 3\cdot \wedge
^{3}\right) \oplus \mathfrak{b}_{12}$ of the little group $SO_{25}$).\medskip

(\ref{coho}) and (\ref{coho-2}) define a cohomological construction of the $%
8 $-dimensional $\mathfrak{e}_{8}$ root lattice and of the $24$-dimensional
Leech lattice $\Lambda _{24}$, respectively based on the analogy between :

\begin{itemize}
\item M-theory in $10+1$ space-time dimensions, with $SO_{9}$ massless
little group and massless spectrum given by $\mathbf{128}$ (gravitino $%
\mathbf{\psi }$) $=$ $\mathbf{84}$ (3-form potential $\wedge ^{3}$)$\oplus
\mathbf{44}$ (graviton $g\simeq S_{2}^{0}$); this corresponds to $D0$-branes
(supergravitons) in BFSS M(atrix) model, carrying $256=128(B)+128(F)$ KK
states \cite{BFSS};

\item the would-be $\mathcal{N}=1$ supergravity in $26+1$ space-time
dimensions, with $SO_{25}$ massless little group and massless spectrum given
by $\mathbf{98,304}$ (would-be gravitino $\mathbf{\psi }$) $=3\cdot \mathbf{%
2,300}\oplus 3\cdot \mathbf{12,650}\oplus \mathbf{53,130}$ (set of massless $%
p$-forms which is the \textquotedblleft ($26+1$)-dimensional
analogue\textquotedblright\ of the 3-form in $10+1$)$\oplus \mathbf{324}$
(graviton $g\simeq S_{2}^{0}$); this would correspond to $D0$-branes (i.e.,
the would-be \textquotedblleft supergravitons\textquotedblright ) in the
would-be BFSS-like M(atrix) model, carrying $196,608=98,304(B)+98,304(F)$ KK
states.
\end{itemize}

There are many analogies, but the big difference is (local) supersymmetry in
$D=26+1$ (and\ possibly in $D=25+1$), whose nature is at present still
conjectural.

The \textquotedblleft Leech algebra\textquotedblright\ $\mathfrak{L}_{24}$
encodes $\dim \mathfrak{su}_{25}=624=324+300$, and $2\cdot 97,980=2\cdot
(3\cdot 2,300+3\cdot 12,650+53,130)=195,960$ to get $624+195,960=196,584$.
Removing the $12+12=24$ Cartans gives $196,560$, which is the number of
minimal Leech vectors. It is thus tempting to conjecture \textquotedblleft
Monstrous supergravitons\textquotedblright\ as $D0$-branes, as $\mathfrak{L}%
_{24}$ \textquotedblleft sees\textquotedblright\ $98,304$ of the bosonic KK
states. On the other hand, the Monster $\mathbb{M}$ acts on almost all of
these, albeit seeing only $299+1$ of the $324$ graviton degrees of freedom
from $324+300$, giving $299+1+(300+97,980)=299+1+98,280$ of the Griess
algebra \cite{griess76,conwaybook}.

Therefore, \textit{the relation between the \textquotedblleft Leech
algebra\textquotedblright\ }$\mathfrak{L}_{24}$\textit{\ and the Griess
algebra is realized in field theory by the relation between M}$^{2}$\textit{%
-theory and its subsector\ (\ref{sub}) coupled to one RS field (the would-be
gravitino) in} $D=26+1$, discussed in Sec. \ref{BBFF}.

\subsection{$26+1\longrightarrow 10+1$ through Vinberg's T-algebras}

How can one relate M-theory in $D=10+1$ with M$^{2}$-theory in $D=26+1$?

The dimensional reduction $26+1\longrightarrow 10+1$ may have a non-trivial
structure: one can proceed along a decomposition proved by Wilson \cite%
{wilsonLeech}, characterizing the aforementioned number of minimal Leech
vectors as%
\begin{equation}
196,560=3\cdot 240\cdot (1+16+256).  \label{pre-HT}
\end{equation}%
Therefore, we identify $1+16+256=273$ with the (Hermitian part of) Vinberg's
T-algebra and $240$ with $E_8$ fibers\footnote{%
In (\ref{HT}) the Greek subscripts discriminate among $\mathfrak{so}_{16}$%
-singlets.} \cite{Vinberg,geoEYM}%
\begin{equation}
T_{3}^{8,2}=\underset{\text{written~in~a~}\mathfrak{so}_{16}~\text{%
covariant~way}}{\left(
\begin{array}{ccc}
\mathbf{1}_{\alpha } & \mathbf{16} & \mathbf{128} \\
\ast & \mathbf{1}_{\beta } & \mathbf{128}^{\prime } \\
\ast & \ast & \mathbf{1}_{\gamma }%
\end{array}%
\right) },  \label{HT}
\end{equation}%
with spin factor lightcone coordinates $\mathbf{1}_{\alpha }$ and $\mathbf{1}%
_{\beta }$ removed, thus yielding $128+128+16+1=273$ degrees of freedom. The
spin factor $\mathbf{1}_{\alpha }\oplus \mathbf{1}_{\beta }\oplus \mathbf{16}
$ of $T_{3}^{8,2}$ (\ref{HT}) enjoys\ an enhancement from $\mathfrak{so}%
_{16} $~(massless little algebra in $17+1$) to $\mathfrak{so}_{17,1}$
Lorentz algebra\footnote{$\mathfrak{so}_{17,1}$ would be the Lorentz
symmetry of the 18-dimensional string theory suggested by Lorentzian
Kac-Moody algebras \cite{lorentzKM}.}, and $\mathfrak{der}\left(
T_{3}^{8,2}\right) =$mcs$\left( \mathfrak{so}_{17,1}\right) =\mathfrak{so}%
_{17}$ \cite{ep3}. Breaking the $\mathfrak{so}_{25}$ Lie algebra of massless
little group in $26+1$ with respect to $\mathfrak{so}_{17}$, as well as its $%
\mathbf{4,096}$ spinor (both encoded in the so-called \textquotedblleft
Exceptional Periodicity\textquotedblright\ algebra $\mathfrak{f}_{4}^{3}$
\cite{MSEP}), one obtains the decomposition%
\begin{equation}
\mathfrak{f}_{4}^{3}:=\mathfrak{so}_{25}\oplus \mathbf{4,096}=\mathfrak{so}%
_{17}\oplus \mathfrak{so}_{8}\oplus (\mathbf{17},\mathbf{8}_{v})\oplus (%
\mathbf{256},\mathbf{8}_{s})\oplus (\mathbf{256},\mathbf{8}_{c}).
\end{equation}%
As $\mathfrak{so}_{8}$ acts on $S^{7}$, one can take the $240$ roots as
forming a discrete 7-sphere, and the $273$ is constructed as $17+256=273$ by
picking one of the $\mathbf{256}$ spinors. This gives a discrete form of the
maximal Hopf fibration%
\begin{equation}
S^{7}\hookrightarrow S^{15}\rightarrow S^{8},
\end{equation}%
and the three maps yield three charts of the form $196,560=3\cdot 240\cdot
273$ (cfr. (\ref{pre-HT})) in a discrete Cayley plane \cite%
{uleech,geoEYM,wmtheory}. Through the super-Ehlers embedding \cite%
{superehlers}%
\begin{equation}
\mathfrak{e}_{8(8)}=\mathfrak{sl}_{9}\left( \mathbb{R}\right) \oplus \mathbf{%
84}\oplus \mathbf{84}^{\prime }=\mathfrak{so}_{9}\oplus \mathbf{44}\oplus
\mathbf{84}\oplus \mathbf{84},
\end{equation}
we can identify each discrete $S^{7}$ fiber of $240$ $E_{8}$ roots with the
M2- and M5- brane gauge fields of $D=10+1$ M-theory, as well as with little
group ($\mathfrak{so}_{9}$) and graviton ($\mathbf{44}$) degrees of freedom,
albeit with all $4+4$ Cartans removed. This is understood with $\mathfrak{so}%
_{9}\subset \mathfrak{so}_{25}$ acting isometrically on the $S^{8}$ base.
From this perspective, the reduction from $D=26+1$ to $D=10+1$ occurs first
along three charts, and gauge and gravity data are encoded in discrete $%
S^{7} $ chart fibers therein.

This picture is further supported by noting that the Conway group $Co_{0}$
is a maximal finite subgroup of $SO_{24}$, and that $Co_{0}$ can be
generated by unitary $3\times 3$ octonionic matrices \cite{wilsonLeech} of $%
F_{4}$ type \cite{uleech}. In general, the stabilizer subgroup of $3\times 3$
unitary matrices over the octonions $\mathbb{O}$ lies in $SO_{9}\subset
F_{4} $ through Peirce decomposition; since there are three independent
primitive idempotents in the exceptional Jordan algebra $J_{3}^{\mathbb{O}}$%
, there are three such embedded copies of $SO_{9}$, providing three charts
for the reduction $26+1\longrightarrow 10+1$.

\subsection{$10+1\longrightarrow 3+1$ through $S^{7}$ fiber}

As it is well known, a remarkable class of M-theory compactifications is
provided by $G_{2}$ compactifications to $D=3+1$, where the internal
manifold with $G_{2}$ holonomy is characterized by its invariant 3-form
(which comes from an octonionic structure) \cite{Gukov}. In the $26+1$
framework under consideration, a compactification down to $3+1$ dimensions
can involve a 23-sphere $S^{23}$, which in turn can be fibrated with an $%
\mathbb{OP}^{2}$ base and $S^{7}$ fibers. Since $S^{7}$ is the
quintessential $G_{2}$ manifold \cite{Joyce}, this provides a natural $%
26+1\longrightarrow 10+1\longrightarrow 3+1$ pattern of reduction along a $%
G_{2}$ manifold from Monstrous M-theory.

\section{Further evidence for M$^{2}$-theory : Monster SCFT and massless $p$%
-forms in $25+1$}

In order to conclude the present investigation of higher-dimensional gravity
theories which can exhibit the Monster group as symmetry of their massless
spectrum, we reconsider Witten's Monster $\mathcal{N}=1$ SCFT dual to
three-dimensional gravity \cite{witten3d}. We will show that the
coefficients of its partition function enjoy rather simple interpretations
as sums of degrees of freedom of massless fields in $D=25+1$ space-time
dimensions, namely as sums of dimensions of suitable representations of the
corresponding massless little group $SO_{24}$. This fact provides further
evidence of how a purely bosonic theory of gravity and\ massless $p$-forms
in $25+1$ space-time dimensions can be probed by the Monster group $\mathbb{M%
}$ in terms of its lowest dimensional representations.

We start and recall the partition function of Witten's $\mathcal{N}=1$
Monster SCFT (cfr. e.g. (3.35) of \cite{witten3d}) :%
\begin{eqnarray}
K\left( q\right) &=&q^{-1/2}+276q^{1/2}+2,048q^{1}+11,202q^{3/2}+49,152q^{2}
\notag \\
&&+184,024q^{5/2}+614,400q^{3}+1,881,471q^{7/2}+\mathcal{O}(q^{4}),
\label{pf} \\
&=&q^{1/2}Z^{2B}\left( q\right) ,  \label{pf-2}
\end{eqnarray}%
where $Z^{2B}(q)$ is the 2B McKay-Thompson series (cfr. e.g. (C.1) of \cite%
{fmonster}). The coefficients of $K\left( q\right) $, which in \cite%
{witten3d} have been related to the (smallest) representations of the
Monster group $\mathbb{M}$ \cite{Reprs-Co_1}, also admit a rather simple (in
generally not unique, especially for large coefficients) interpretation in
terms of representations of $SO_{24}$, thus strengthening the evidence for
the existence a gravitational field theory probed by the lowest-dimensional,
non trivial representation(s) of $\mathbb{M}$ itself. Indeed, a tedious but
straightforward computation yields to the following result :%
\begin{equation}
\begin{array}{l}
276~=~\underset{\mathbf{276}}{\left\vert \wedge ^{2}\right\vert }; \\
2,048~=~\underset{\mathbf{2,048}}{\left\vert \mathbf{\lambda }\right\vert };
\\
11,202=~\underset{\mathbf{24}}{\left\vert \wedge ^{1}\right\vert }+2\underset%
{\mathbf{276}}{\left\vert \wedge ^{2}\right\vert }+\underset{\mathbf{10,626}}%
{\left\vert \wedge ^{4}\right\vert }; \\
49,152=\underset{\mathbf{2,048}}{\left\vert \mathbf{\lambda }\right\vert }+%
\underset{\mathbf{47,104}}{\left\vert \mathbf{\psi }\right\vert }; \\
184,024=~~\underset{\mathbf{2,048}}{\left\vert \mathbf{\lambda }\right\vert }%
+\underset{\mathbf{47,104}}{\left\vert \mathbf{\psi }\right\vert }+\underset{%
\mathbf{276}}{\left\vert \wedge ^{2}\right\vert }+\underset{\mathbf{134,596}}%
{\left\vert \wedge ^{6}\right\vert }; \\
614,400=~2\underset{\mathbf{2,048}}{\left\vert \mathbf{\lambda }\right\vert }%
+2\underset{\mathbf{47,104}}{\left\vert \mathbf{\psi }\right\vert }+\underset%
{\mathbf{516,096}}{\left\vert \mathbf{\psi }^{(2)}\right\vert }; \\
1,881,471=~23\underset{\mathbf{1}}{\left\vert \phi \right\vert }+\underset{%
\mathbf{24}}{2\left\vert \wedge ^{1}\right\vert }+4\underset{\mathbf{276}}{%
\left\vert \wedge ^{2}\right\vert }+\underset{\mathbf{2,024}}{\left\vert
\wedge ^{3}\right\vert }+\underset{\mathbf{10,626}}{\left\vert \wedge
^{4}\right\vert }+3\underset{\mathbf{42,504}}{\left\vert \wedge
^{5}\right\vert }+2\underset{\mathbf{134,596}}{\left\vert \wedge
^{6}\right\vert }+2\underset{\mathbf{735,471}}{\left\vert \wedge
^{8}\right\vert },%
\end{array}
\label{dec-1}
\end{equation}%
where $\mathbf{\psi }^{(p)}$ denotes the $p$-form spinor representation of $%
SO_{24}$, and we have used the notation $\mathbf{\psi }^{(0)}\equiv \mathbf{%
\lambda }$, $\mathbf{\psi }^{(1)}\equiv \mathbf{\psi }$ (cfr. Sec. \ref%
{triality}).

Remarkably, \textit{the degrees of freedom of }$p$\textit{-form spinors} $%
\mathbf{\psi }^{(p)}$ \textit{can} \textit{always} \textit{be expressed}
\textit{only} \textit{in terms of the degrees of freedom of }$p$\textit{%
-form fields} : for the first cases, i.e. for $p=0$, $1$ and $2$, by
recalling $\mathbf{\lambda }$-triality (\ref{1}) (which in turn implies $%
\mathbf{\psi }$-triality (\ref{2})), it holds that%
\begin{eqnarray}
\underset{\text{(}\mathbf{\lambda }\text{-triality (\ref{1}))}}{p=0} &:&~%
\underset{\mathbf{2,048}}{\left\vert \mathbf{\psi }^{(0)}\right\vert }=%
\underset{\mathbf{24}}{\left\vert \wedge ^{1}\right\vert }+\underset{\mathbf{%
2,024}}{\left\vert \wedge ^{3}\right\vert };  \label{pt} \\
\underset{\text{(}\mathbf{\psi }\text{-triality (\ref{2}))}}{p=1} &:&%
\underset{\mathbf{47,104}}{\left\vert \mathbf{\psi }^{(1)}\right\vert }=2%
\underset{\mathbf{276}}{\left\vert \wedge ^{2}\right\vert }+2\underset{%
\mathbf{2,024}}{\left\vert \wedge ^{3}\right\vert }+4\underset{\mathbf{10,626%
}}{\left\vert \wedge ^{4}\right\vert }=2\underset{\mathbf{276}}{\left\vert
\wedge ^{2}\right\vert }+2\underset{\mathbf{2,024}}{\left\vert \wedge
^{3}\right\vert }+\underset{\mathbf{42,504}}{\left\vert \wedge
^{5}\right\vert };  \label{ptt} \\
\underset{\text{(}\mathbf{\psi }^{(2)}\text{-triality)}}{p=2} &:&\underset{%
\mathbf{516,096}}{\left\vert \mathbf{\psi }^{(2)}\right\vert }=14\underset{%
\mathbf{1}}{\left\vert \phi \right\vert }+8\underset{\mathbf{24}}{\left\vert
\wedge ^{1}\right\vert }+\underset{\mathbf{276}}{\left\vert \wedge
^{2}\right\vert }+5\underset{\mathbf{2,024}}{\left\vert \wedge
^{3}\right\vert }+3\underset{\mathbf{10,626}}{\left\vert \wedge
^{4}\right\vert }+3\underset{\mathbf{42,504}}{\left\vert \wedge
^{5}\right\vert }+\underset{\mathbf{346,104}}{\left\vert \wedge
^{7}\right\vert }.  \label{terza}
\end{eqnarray}%
Thus, by using (\ref{pt})-(\ref{terza}), the sums on the r.h.s.'s of (\ref%
{dec-1}) can be expressed only in terms of $p$-form bosonic fields, as
follows :
\begin{equation}
\begin{array}{l}
276~=~\underset{\mathbf{276}}{\left\vert \wedge ^{2}\right\vert }; \\
2,048~=~~\underset{\mathbf{24}}{\left\vert \wedge ^{1}\right\vert }+\underset%
{\mathbf{2,024}}{\left\vert \wedge ^{3}\right\vert }; \\
11,202=~\underset{\mathbf{24}}{\left\vert \wedge ^{1}\right\vert }+2\underset%
{\mathbf{276}}{\left\vert \wedge ^{2}\right\vert }+\underset{\mathbf{10,626}}%
{\left\vert \wedge ^{4}\right\vert }; \\
49,152=~\underset{\mathbf{24}}{\left\vert \wedge ^{1}\right\vert }+2\underset%
{\mathbf{276}}{\left\vert \wedge ^{2}\right\vert }+3\underset{\mathbf{2,024}}%
{\left\vert \wedge ^{3}\right\vert }+\underset{\mathbf{42,504}}{\left\vert
\wedge ^{5}\right\vert }; \\
184,024=~\underset{\mathbf{24}}{\left\vert \wedge ^{1}\right\vert }+3%
\underset{\mathbf{276}}{\left\vert \wedge ^{2}\right\vert }+3\underset{%
\mathbf{2,024}}{\left\vert \wedge ^{3}\right\vert }+\underset{\mathbf{42,504}%
}{\left\vert \wedge ^{5}\right\vert }+\underset{\mathbf{134,596}}{\left\vert
\wedge ^{6}\right\vert }; \\
614,400=~16\underset{\mathbf{1}}{\left\vert \phi \right\vert }+8\underset{%
\mathbf{24}}{\left\vert \wedge ^{1}\right\vert }+7\underset{\mathbf{276}}{%
\left\vert \wedge ^{2}\right\vert }+2\underset{\mathbf{2,024}}{\left\vert
\wedge ^{3}\right\vert }+3\underset{\mathbf{42,504}}{\left\vert \wedge
^{5}\right\vert }+\underset{\mathbf{134,596}}{\left\vert \wedge
^{6}\right\vert }+\underset{\mathbf{346,104}}{\left\vert \wedge
^{7}\right\vert }; \\
1,881,471=~23\underset{\mathbf{1}}{\left\vert \phi \right\vert }+\underset{%
\mathbf{24}}{2\left\vert \wedge ^{1}\right\vert }+4\underset{\mathbf{276}}{%
\left\vert \wedge ^{2}\right\vert }+\underset{\mathbf{2,024}}{\left\vert
\wedge ^{3}\right\vert }+\underset{\mathbf{10,626}}{\left\vert \wedge
^{4}\right\vert }+3\underset{\mathbf{42,504}}{\left\vert \wedge
^{5}\right\vert }+2\underset{\mathbf{134,596}}{\left\vert \wedge
^{6}\right\vert }+2\underset{\mathbf{735,471}}{\left\vert \wedge
^{8}\right\vert }.%
\end{array}
\label{dec-2}
\end{equation}

Thus, the first coefficients of the partition function (\ref{pf})-(\ref{pf-2}%
) of the $\mathcal{N}=1$ Monster SCFT \cite{witten3d} can be decomposed as
sums of the dimensions of purely bosonic, $p$-form representations of $%
SO_{24}$; since this latter is the massless little group in $D=25+1$
space-time dimensions, the above results imply that, \textit{at least} for
the first coefficients, \textit{the coefficients of the partition function
of }$\mathcal{N}=1$ \textit{Monster SCFT can be expressed in terms of
degrees of freedom of massless, purely bosonic,} $p$\textit{-form fields in }%
$25+1$\textit{\ space-time dimensions}.

The purely bosonic nature of such degrees of freedom is ultimately due to
the $\mathbf{\lambda }$-triality (\ref{1}) (or, equivalently, (\ref{pt})),
which is the generalization of the triality $\mathbb{T}$, discussed at the
start of Sec. \ref{triality} from $8$ to $24$ dimensions. To the best of our
knowledge, no other examples of such a generalized, \textquotedblleft
weak\textquotedblright\ triality are known in other dimensions, so $24$
stands out as a very peculiar number in this respect.

Note how all the purely bosonic decompositions (\ref{dec-2}) share a common
feature: for each $p\geqslant 0$, the decompositions (\ref{dec-2}) exhibit
the lowest possible multiplicity of $p$-form fields, constrained to
correspond to a number of degrees of freedom which is strictly smaller than
the dimensions of the subsequent ($p+1$)-form field: namely, the condition%
\begin{equation}
\#\wedge ^{p}\cdot \left\vert \wedge ^{p}\right\vert \leqslant \left\vert
\wedge ^{p+1}\right\vert
\end{equation}%
holds in (\ref{dec-2}) for all $p$'s appearing.

\section{\label{Conclusion}Final remarks}

\textbf{Monstrous M-theory, Monstrous dilatonic gravities and Monstrous
Moonshine}

We have shown that in $26+1$ space-time dimensions there exists a Monstrous
M-theory, or simply M$^{2}$-theory, whose massless spectrum (\ref{fieldc'})
contains $196,884$ degrees of freedom that may be acted upon by the Monster
group $\mathbb{M}$ after reduction to $D=25+1$, because it corresponds to
the sum of the two smallest representations of $\mathbb{M}$, namely the
trivial (singlet) $\mathbf{1}$ and the non-trivial one $\mathbf{196,883}$. A
subsector of M$^{2} $-theory yields Horowitz and Susskind's bosonic M-theory
\cite{bMtheory}. Crucially, the disentangling of the $196,884$ degrees of
freedom into $\mathbf{196,883}\oplus \mathbf{1}$ occurs only when reducing M$%
^{2}$-theory down to $25+1$, obtaining the massless spectrum (\ref{fieldc}),
in which the dilaton $\phi $ is identified with the singlet of $\mathbb{M}$
: in other words, \textit{the (initial observation giving rise to) Monstrous
Moonshine} \cite{conway79}\textit{\ is crucially related to the KK
compactification of M}$^{2}$\textit{-theory down to a certain Monstrous
dilatonic gravity\footnote{%
Namely, the theory $0.\alpha .iii$ within the classification carried out in
Sec. \ref{class}.} in }$25+1$\textit{\ space-time dimensions}.

Remarkably, such a Monstrous dilatonic theory in $25+1$ contains a subsector
given by the massless excitations of the closed and open bosonic string in $%
25+1$, namely a graviton, an antisymmetric rank-2 field, a dilaton, and a
1-form potential. Actually, by generalizing the \textit{triality} $\mathbb{T}
$ of $SO_{8}$ (massless little group of string theory in $9+1$) to $SO_{24}$
(massless little group of bosonic string theory in $25+1$), such a dilatonic
(Einstein) gravity theory can be shown to be part of a web of some $60$
gravito-dilatonic theories, collectively named \textit{Monstrous gravity
theories}, whose coarse-grained classification is given in Sec. \ref{class}.

The relation between $SO_{8}$ and $SO_{24}$ (which at present is the unique
dimension enjoying a kind of generalization of $\mathbb{T}$) can be
interpreted in terms of the Conway group\footnote{%
The Conway group $Co_{0}$ is the full automorphism of the Leech lattice $%
\Lambda _{24}$; however, it is not a simple group, nor is it contained in
the Monster. In fact, its quotient by its center $\mathbb{Z}_{2}$, namely
the Conway simple group $Co_{1}\sim Co_{0}/\mathbb{Z}_{2}$ is contained in $%
\mathbb{M}$. This means the Monster's maximal finite subgroup $Co_{1}$ has
the $\mathbb{Z}_{2}$ action built in, which acts on only half the minimal
Leech vectors $196,560/2=98,280$.} $Co_{0}$, which is a maximal finite
subgroup of $SO_{24}$ itself; as shown by Wilson \cite{wilsonLeech}, $Co_{0}$
is generated by unitary $3\times 3$ octonion matrices, namely by $F_{4}$
matrices \cite{uleech}. Interestingly, $SO_{9}$ can be maximally embedded
into $F_{4}$ in three possible ways, each one providing the manifestly $%
\mathbb{T}$-invariant breaking%
\begin{equation}
\mathfrak{f}_{4}\rightarrow \mathfrak{so}_{9}=\mathfrak{so}_{8}\oplus
\mathbf{8}_{v}\oplus \mathbf{8}_{s}\oplus \mathbf{8}_{c};
\end{equation}%
in this sense, no triality is needed for $\mathfrak{so}_{24}$, but rather
just the threefold nature of the (symmetric) embedding $SO_{9}\subset F_{4}$%
. In turn, the \textquotedblleft anomalous" embedding \cite{Ramond}%
\begin{equation}
\mathfrak{f}_{4}\oplus \mathbf{273}\hookrightarrow \mathfrak{so}_{26}
\end{equation}
allows one to reduce from $26+1$ to lower dimensions in a non-trivial way,
namely along the chain $26+1\rightarrow 25+1\rightarrow 10+1\rightarrow 3+1$%
. this, as remarked in \cite{wmtheory}, confirms and strengthens Ramond's
and Sati's argument that $D=10+1$ M-theory has hidden Cayley plane $\mathbb{O%
}\mathbb{P}^{2}$ fibers \cite{Sati}.

The Moonshine decomposition (\ref{deccc}),%
\begin{equation}
196,884=\mathbf{196,883}\oplus \mathbf{1}
\end{equation}%
always holds in Monstrous gravities, due to the very existence of the
dilatonic scalar field $\phi $ in their spectrum. In particular, the dilaton
$\phi $ is a singlet of $\mathbb{M}$. Monstrous gravities in $25+1$
space-time dimensions, and the presence of a unique $\phi $, are intimately
related to the representation $\mathbf{196,883}$ of $\mathbb{M}$, and thus
\textit{they may provide an explanation of the (initial observation giving
rise to) Monstrous Moonshine in terms of (higher-dimensional, gravitational)
field theory}.\medskip

\textbf{Black hole entropy in }$\mathbf{2+1}$

Along the lines of Witten's invetsigation of three-dimensional gravity \cite%
{witten3d}, the present paper suggests that \textit{the quantum entropy} $%
\ln \left( 196,883\right) \simeq 12.19$ \textit{has a manifest
higher-dimensional interpretation}, since the BTZ black hole degrees of
freedom can be expressed in terms of massless degrees of freedom of fields
in $25+1$ space-time dimensions.\medskip

\textbf{Local SUSY in }$\mathbf{26+1}$\textbf{\ ?}

Remarkably, a certain subsector of the spectrum of M$^{2}$-theory, given by (%
\ref{sub}), when coupled to one massless Rarita-Schwinger field $\psi $ in $%
26+1$, gives rise to a theory which has the same number of bosonic and
fermionic massless degrees of freedom, namely%
\begin{equation}
B=F=98,304,
\end{equation}%
for a total of $196,608$ degrees of freedom. We have been therefore tempted
to ask ourselves to ask whether this subsector of M$^{2}$-theory, when
coupled to a RS field $\psi $, may actually enjoy (local) supersymmetry in $%
26+1$ space-time dimensions, thus giving rise to a would-be $\mathcal{N}=1$,
$D=26+1$ supergravity theory. In this line of reasoning, we have conjectured
a \textquotedblleft M-theory-inspired\textquotedblright\ Lagrangian density,
as well as the corresponding local supersymmetry transformations in $26+1$.
The invariance of such a Lagrangian under those supersymmetry
transformations is still conjectural, and to prove (or disprove) it seems
quite a formidable, though absolutely worthy task, and we leave it for
further future work.

At any rate, the reduction of the bosonic sector (\ref{sub}) of such a
would-be $\mathcal{N}=1$ supergravity from $26+1$ to $25+1$ yields a
suitable subsector of the Monstrous gravity labelled by $2.\gamma .ii$ in
the classification of Sec. \ref{class}, simply obtained by letting $\#\wedge
^{2}:4\longrightarrow 3$. In light of this, we cannot help but point out a
certain mismatch, essentially amounting to the $\mathbf{276}$ degrees of
freedom of a massless 2-form in $25+1$, between the total (bosonic $+$
fermionic) degrees of freedom of the would-be $\mathcal{N}=1$ supergravity
in $26+1$ ($98,304+98,304=196,608$) and the (purely bosonic) $196,884$
degrees of freedom of M$^{2}$-theory : $196,884-196,608=276$. In this sense,
\textquotedblleft monstrousity\textquotedblright\ and (would-be)
\textquotedblleft supersymmetry\textquotedblright\ in $26+1$ (as well as,
predictably, in $25+1$) space-time dimensions exhibit a slight disalignment,
though being tightly related.\medskip

\textbf{Leech lattice and Griess algebra }

All this suggests that the Monster group $\mathbb{M}$ has its origin in a
gravity theory in $26+1$ dimensions, as its definition as the automorphism
of the Griess algebra \cite{griess76,flm,borcherds2002} is clarified by
showing that such an algebra is not merely a sum of unrelated spaces, but
related to the massless spectrum of Monstrous gravities in $25+1$, which in
\textit{at least} one case (namely, the $0.\alpha .iii$ theory, whose
massless spectrum is given by (\ref{pre-fieldc})-(\ref{fieldc})) oxidates up
to M$^{2}$-theory in $26+1$. The spectrum of M$^{2}$-theory dimensionally
reduced to $25+1$ contains a subsector given by the massless excitations of
the closed and open bosonic string in $25+1$, namely a graviton, an
antisymmetric rank-2 field, a dilaton, and a 1-form potential. Therefore,
\textit{the relation between the \textquotedblleft Leech
algebra\textquotedblright\ }$\mathfrak{L}_{24}$\textit{\ and the Griess
algebra is realized in field theory by the relation between M}$^{2}$\textit{%
-theory and its subsector\ (\ref{sub}) coupled to one RS field (the would-be
gravitino) in} $26+1$, discussed in Sec. \ref{BBFF}.

On the other hand, the discussion of the analogies between the $\mathfrak{e}%
_{8}$ root lattice and the Leech lattice $\Lambda _{24}$ seems to suggest
that M-theory in $10+1$ and the would-be $\mathcal{N}=1$ supergravity in $%
26+1$ are tightly related to the lattices $\mathfrak{e}_{8}$ resp. $\Lambda
_{24}$, which determine the optimal lattice packings in $D=8$ resp. $24$.

\bigskip

\textbf{Developments}

Many directions for further future developments stem from the present work,
which is a preliminary investigations of higher-dimensional structures in
space-time which reflects themselves in large-dimensional, yet finite, group
theoretical structures. Below, we list a few possible developments.

\begin{itemize}
\item It would be interesting to explore the implications of the
characterization of the $\mathbb{M}$ as acting on the whole massless
spectrum of M$^{2}$-theory in $26+1$ space-time dimensions.

\item One could further study the maps discussed in Sec. \ref{triality}; as
pointed out above, no other Dynkin diagram (besides $\mathfrak{d}_{4}$) has
an automorphism group of order greater than $2$, thus such maps cannot be
realized as an automorphism of $\mathfrak{d}_{12}$, nor they can be traced
back to some structural symmetry of the Dynkin diagram of $\mathfrak{d}_{12}$
itself.

\item Also, one could study the Lagrangian structure of M$^{2}$-theory, as
well as of its Scherk-Schwarz reduction to $25+1$.

\item Further evidence may be gained by investigating whether the dimensions
of representations of finite groups like the Baby Monster group $\mathbb{BM}$%
, the Conway group $Co_{0}$ and the simple Conway group $Co_{1}\simeq Co_{0}/%
\mathbb{Z}_{2}$, can all be rather simply interpreted as sums of dimensions
of representation of $SO_{24}$ or $SO_{25}$ itself, and study the
decomposition of the (smallest) coefficients of the partition functions of
the SCFT's derived from the Monster SCFT.

\item Further study may concern the double copy structure of Monster
dilatonic gravities in $25+1$, as well as of M$^{2}$-theory, and its
possibly supersymmetric subsector, in $26+1$.

\item The investigation on the existence of local SUSY in $26+1$, and the
determination of the corresponding Lagrangian and SUSY transformations is of
utmost relevance, of course.

\item Last but not least, it would be interesting to study the massive
spectrum of (massive variants of) Monstrous gravities and of M$^{2}$-theory.
\end{itemize}

\bigskip

We would like to conclude with a sentence by John H. Conway, to whom this
paper is dedicated, on the Monster group \cite{Conway-cit}:
\textquotedblleft \textit{There's never been any kind of explanation of why
it's there, and it's obviously not there just by coincidence. It's got too
many intriguing properties for it all to be just an accident}."

\section*{Acknowledgements}

We would like to acknowledge an informal yet inspiring discussion with Eric
Weinstein during the \textquotedblleft Advances in Quantum
Gravity\textquotedblright\ Conference, hosted by Laura Deming at Topos
House, San Francisco, CA, on July 2016. Since that occasion, we started
thinking about the idea of a \textquotedblleft weak\textquotedblright\
triality for $\mathfrak{d}_{12}$. Also, we thank Richard Borcherds for
insights on the domain of the fake Monster automorphic form in signature $%
D=26+2$.

\appendix

\section{Chern-Simons Lagrangian terms for Monstrous M-theory}

\label{monsterCS}

The full Lagrangian from Eq.~\eqref{L''} is given by
\begin{eqnarray}
\sqrt{|g|}\mathcal{L}_{CS} &=&\epsilon \mathcal{E}%
_{2}^{ABCiDE}G_{A}^{(6)}G_{B}^{(6)}G_{C}^{(6)}G_{i}^{(4)}G_{D}^{(3)}C_{E}^{(2)}+\epsilon
\mathcal{F}%
_{2}^{ABijCD}G_{A}^{(6)}G_{B}^{(6)}G_{i}^{(5)}G_{j}^{(5)}G_{C}^{(3)}C_{D}^{(2)}
\notag \\
&&+\epsilon \mathcal{G}%
_{2}^{ABijkC}G_{A}^{(6)}G_{B}^{(6)}G_{i}^{(5)}G_{j}^{(4)}G_{k}^{(4)}C_{C}^{(2)}+\epsilon
\mathcal{H}%
_{2}^{AijklB}G_{A}^{(6)}G_{i}^{(5)}G_{j}^{(5)}G_{k}^{(5)}G_{l}^{(4)}C_{B}^{(2)}
\notag \\
&&+\epsilon \mathcal{I}%
_{2}^{ijklmA}G_{i}^{(5)}G_{j}^{(5)}G_{k}^{(5)}G_{l}^{(5)}G_{m}^{(5)}C_{A}^{(2)}+\epsilon
\mathcal{J}%
_{2}^{ABiCDEF}G_{A}^{(6)}G_{B}^{(6)}G_{i}^{(4)}G_{C}^{(3)}G_{D}^{(3)}G_{E}^{(3)}C_{F}^{(2)}
\notag \\
&&+\epsilon \mathcal{K}%
_{2}^{AijBCDE}G_{A}^{(6)}G_{i}^{(5)}G_{j}^{(5)}G_{B}^{(3)}G_{C}^{(3)}G_{D}^{(3)}C_{E}^{(2)}+\epsilon
\mathcal{L}%
_{2}^{AijkBCD}G_{A}^{(6)}G_{i}^{(5)}G_{j}^{(4)}G_{k}^{(4)}G_{B}^{(3)}G_{C}^{(3)}C_{D}^{(2)}
\notag \\
&&+\epsilon \mathcal{M}%
_{2}^{ijklABC}G_{i}^{(5)}G_{j}^{(5)}G_{k}^{(5)}G_{l}^{(4)}G_{A}^{(3)}G_{B}^{(3)}C_{C}^{(2)}+\epsilon
\mathcal{N}%
_{2}^{AijklBC}G_{A}^{(6)}G_{i}^{(4)}G_{j}^{(4)}G_{k}^{(4)}G_{l}^{(4)}G_{B}^{(3)}C_{C}^{(2)}
\notag \\
&&+\epsilon \mathcal{P}%
_{2}^{ijklmAB}G_{i}^{(5)}G_{j}^{(5)}G_{k}^{(4)}G_{l}^{(4)}G_{m}^{(4)}G_{A}^{(3)}C_{B}^{(2)}+\epsilon
\mathcal{W}%
_{2}^{ijklmnA}G_{i}^{(5)}G_{j}^{(4)}G_{k}^{(4)}G_{l}^{(4)}G_{m}^{(4)}G_{n}^{(4)}C_{A}^{(2)}
\notag \\
&&+\epsilon \mathcal{Q}%
_{2}^{AiBCDEFG}G_{A}^{(6)}G_{i}^{(4)}G_{B}^{(3)}G_{C}^{(3)}G_{D}^{(3)}G_{E}^{(3)}G_{F}^{(3)}C_{G}^{(2)}+\epsilon
\mathcal{R}%
_{2}^{ijABCDEF}G_{i}^{(5)}G_{j}^{(5)}G_{A}^{(3)}G_{B}^{(3)}G_{C}^{(3)}G_{D}^{(3)}G_{E}^{(3)}C_{F}^{(2)}
\notag \\
&&+\epsilon \mathcal{S}%
_{2}^{ijkABCDE}G_{i}^{(5)}G_{j}^{(4)}G_{k}^{(4)}G_{A}^{(3)}G_{B}^{(3)}G_{C}^{(3)}G_{D}^{(3)}C_{E}^{(2)}+\epsilon
\mathcal{T}%
_{2}^{ijklABCD}G_{i}^{(4)}G_{j}^{(4)}G_{k}^{(4)}G_{l}^{(4)}G_{A}^{(3)}G_{B}^{(3)}G_{C}^{(3)}C_{D}^{(2)}
\notag \\
&&+\epsilon \mathcal{U}%
_{2}^{iABCDEFGH}G_{i}^{(4)}G_{A}^{(3)}G_{B}^{(3)}G_{C}^{(3)}G_{D}^{(3)}G_{E}^{(3)}G_{F}^{(3)}G_{G}^{(3)}C_{H}^{(2)}+\epsilon
\mathcal{E}%
_{3}^{ABCDi}G_{A}^{(6)}G_{B}^{(6)}G_{C}^{(6)}G_{D}^{(6)}C_{i}^{(3)}  \notag
\\
&&+\epsilon \mathcal{F}%
_{3}^{ABCDEi}G_{A}^{(6)}G_{B}^{(6)}G_{C}^{(6)}G_{D}^{(3)}G_{E}^{(3)}C_{i}^{(3)}+\epsilon
\mathcal{G}%
_{3}^{ABijCk}G_{A}^{(6)}G_{B}^{(6)}G_{i}^{(5)}G_{j}^{(4)}G_{C}^{(3)}C_{k}^{(3)}
\notag \\
&&+\epsilon \mathcal{H}%
_{3}^{AijkBl}G_{A}^{(6)}G_{i}^{(5)}G_{j}^{(5)}G_{k}^{(5)}G_{B}^{(3)}C_{l}^{(3)}+\epsilon
\mathcal{I}%
_{3}^{ABijkl}G_{A}^{(6)}G_{B}^{(6)}G_{i}^{(4)}G_{j}^{(4)}G_{k}^{(4)}C_{l}^{(3)}
\notag \\
&&+\epsilon \mathcal{J}%
_{3}^{Aijklm}G_{A}^{(6)}G_{i}^{(5)}G_{j}^{(5)}G_{k}^{(4)}G_{l}^{(4)}C_{m}^{(3)}+\epsilon
\mathcal{K}%
_{3}^{ijklmn}G_{i}^{(5)}G_{j}^{(5)}G_{k}^{(5)}G_{l}^{(5)}G_{m}^{(4)}C_{n}^{(3)}
\notag \\
&&+\epsilon \mathcal{L}%
_{3}^{ABCDEFi}G_{A}^{(6)}G_{B}^{(6)}G_{C}^{(3)}G_{D}^{(3)}G_{E}^{(3)}G_{F}^{(3)}C_{i}^{(3)}+\epsilon
\mathcal{M}%
_{3}^{AijBCDk}G_{A}^{(6)}G_{i}^{(5)}G_{j}^{(4)}G_{B}^{(3)}G_{C}^{(3)}G_{D}^{(3)}C_{k}^{(3)}
\notag \\
&&+\epsilon \mathcal{N}%
_{3}^{ijkABCl}G_{i}^{(5)}G_{j}^{(5)}G_{k}^{(5)}G_{A}^{(3)}G_{B}^{(3)}G_{C}^{(3)}C_{l}^{(3)}+\epsilon
\mathcal{O}%
_{3}^{AijkBCl}G_{A}^{(6)}G_{i}^{(4)}G_{j}^{(4)}G_{k}^{(4)}G_{B}^{(3)}G_{C}^{(3)}C_{l}^{(3)}
\notag \\
&&+\epsilon \mathcal{P}%
_{3}^{ijklABm}G_{i}^{(5)}G_{j}^{(5)}G_{k}^{(4)}G_{l}^{(4)}G_{A}^{(3)}G_{B}^{(3)}C_{m}^{(3)}+\epsilon
\mathcal{R}%
_{3}^{ijklmAn}G_{i}^{(5)}G_{j}^{(4)}G_{k}^{(4)}G_{l}^{(4)}G_{m}^{(4)}G_{A}^{(3)}C_{n}^{(3)}
\\
&&+\epsilon \mathcal{S}%
_{3}^{ijklmno}G_{i}^{(4)}G_{j}^{(4)}G_{k}^{(4)}G_{l}^{(4)}G_{m}^{(4)}G_{n}^{(4)}C_{o}^{(3)}+\epsilon
\mathcal{T}%
_{3}^{ABCDEFGi}G_{A}^{(6)}G_{B}^{(3)}G_{C}^{(3)}G_{D}^{(3)}G_{E}^{(3)}G_{F}^{(3)}G_{G}^{(3)}C_{i}^{(3)}
\notag \\
&&+\epsilon \mathcal{U}%
_{3}^{ijABCDEk}G_{i}^{(5)}G_{j}^{(4)}G_{A}^{(3)}G_{B}^{(3)}G_{C}^{(3)}G_{D}^{(3)}G_{E}^{(3)}C_{k}^{(3)}+\epsilon
\mathcal{V}%
_{3}^{ijkABCDl}G_{i}^{(4)}G_{j}^{(4)}G_{k}^{(4)}G_{A}^{(3)}G_{B}^{(3)}G_{C}^{(3)}G_{D}^{(3)}C_{l}^{(3)}
\notag \\
&&+\epsilon \mathcal{W}%
_{3}^{ABCDEFGHi}G_{A}^{(3)}G_{B}^{(3)}G_{C}^{(3)}G_{D}^{(3)}G_{E}^{(3)}G_{F}^{(3)}G_{G}^{(3)}G_{H}^{(3)}C_{i}^{(3)}+\epsilon
\mathcal{E}%
_{4}^{ABCij}G_{A}^{(6)}G_{B}^{(6)}G_{C}^{(6)}G_{i}^{(5)}C_{j}^{(4)}  \notag
\\
&&+\epsilon \mathcal{F}%
_{4}^{ABiCDj}G_{A}^{(6)}G_{B}^{(6)}G_{i}^{(5)}G_{C}^{(3)}G_{D}^{(3)}C_{j}^{(4)}+\epsilon
\mathcal{G}%
_{4}^{ABijCk}G_{A}^{(6)}G_{B}^{(6)}G_{i}^{(4)}G_{j}^{(4)}G_{C}^{(3)}C_{k}^{(4)}
\notag \\
&&+\epsilon \mathcal{H}%
_{4}^{AijkBl}G_{A}^{(6)}G_{i}^{(5)}G_{j}^{(5)}G_{k}^{(4)}G_{B}^{(3)}C_{l}^{(4)}+\epsilon
\mathcal{I}%
_{4}^{ijklAm}G_{i}^{(5)}G_{j}^{(5)}G_{k}^{(5)}G_{l}^{(5)}G_{A}^{(3)}C_{m}^{(4)}
\notag \\
&&+\epsilon \mathcal{J}%
_{4}^{Aijklm}G_{A}^{(6)}G_{i}^{(5)}G_{j}^{(4)}G_{k}^{(4)}G_{l}^{(4)}C_{m}^{(4)}+\epsilon
\mathcal{K}%
_{4}^{ijklmn}G_{i}^{(5)}G_{j}^{(5)}G_{k}^{(5)}G_{l}^{(4)}G_{m}^{(4)}C_{n}^{(4)}
\notag \\
&&+\epsilon \mathcal{L}%
_{4}^{AiBCDEj}G_{A}^{(6)}G_{i}^{(5)}G_{B}^{(3)}G_{C}^{(3)}G_{D}^{(3)}G_{E}^{(3)}C_{j}^{(4)}+\epsilon
\mathcal{M}%
_{4}^{AijBCDk}G_{A}^{(6)}G_{i}^{(4)}G_{j}^{(4)}G_{B}^{(3)}G_{C}^{(3)}G_{D}^{(3)}C_{k}^{(4)}
\notag \\
&&+\epsilon \mathcal{N}%
_{4}^{ijkABCl}G_{i}^{(5)}G_{j}^{(5)}G_{k}^{(4)}G_{A}^{(3)}G_{B}^{(3)}G_{C}^{(3)}C_{l}^{(4)}+\epsilon
\mathcal{O}%
_{4}^{ijklABm}G_{i}^{(5)}G_{j}^{(4)}G_{k}^{(4)}G_{l}^{(4)}G_{A}^{(3)}G_{B}^{(3)}C_{m}^{(4)}
\notag \\
&&+\epsilon \mathcal{P}%
_{4}^{ijklmAn}G_{i}^{(4)}G_{j}^{(4)}G_{k}^{(4)}G_{l}^{(4)}G_{m}^{(4)}G_{A}^{(3)}C_{n}^{(4)}+\epsilon
\mathcal{Q}%
_{4}^{iABCDEFj}G_{i}^{(5)}G_{A}^{(3)}G_{B}^{(3)}G_{C}^{(3)}G_{D}^{(3)}G_{E}^{(3)}G_{F}^{(3)}C_{j}^{(4)}
\notag \\
&&+\epsilon \mathcal{R}%
_{4}^{ijABCDEk}G_{i}^{(4)}G_{j}^{(4)}G_{A}^{(3)}G_{B}^{(3)}G_{C}^{(3)}G_{D}^{(3)}G_{E}^{(3)}C_{k}^{(4)}+\epsilon
\mathcal{E}%
_{5}^{ABCiD}G_{A}^{(6)}G_{B}^{(6)}G_{C}^{(6)}G_{i}^{(4)}C_{D}^{(5)}  \notag
\\
&&+\epsilon \mathcal{F}%
_{5}^{ABijC}G_{A}^{(6)}G_{B}^{(6)}G_{i}^{(5)}G_{j}^{(5)}C_{C}^{(5)}+\epsilon
\mathcal{G}%
_{5}^{ABiCDE}G_{A}^{(6)}G_{B}^{(6)}G_{i}^{(4)}G_{C}^{(3)}G_{D}^{(3)}C_{E}^{(5)}
\notag \\
&&+\epsilon \mathcal{H}%
_{5}^{AijBCD}G_{A}^{(6)}G_{i}^{(5)}G_{j}^{(5)}G_{B}^{(3)}G_{C}^{(3)}C_{D}^{(5)}+\epsilon
\mathcal{I}%
_{5}^{AijkBC}G_{A}^{(6)}G_{i}^{(5)}G_{j}^{(4)}G_{k}^{(4)}G_{B}^{(3)}C_{C}^{(5)}
\notag \\
&&+\epsilon \mathcal{J}%
_{5}^{ijklAB}G_{i}^{(5)}G_{j}^{(5)}G_{k}^{(5)}G_{l}^{(4)}G_{A}^{(3)}C_{B}^{(5)}+\epsilon
\mathcal{K}%
_{5}^{AijklB}G_{A}^{(6)}G_{i}^{(4)}G_{j}^{(4)}G_{k}^{(4)}G_{l}^{(4)}C_{B}^{(5)}
\notag \\
&&+\epsilon \mathcal{L}%
_{5}^{ijklmA}G_{i}^{(5)}G_{j}^{(5)}G_{k}^{(4)}G_{l}^{(4)}G_{m}^{(4)}C_{A}^{(5)}+\epsilon
\mathcal{M}%
_{5}^{AiBCDEF}G_{A}^{(6)}G_{i}^{(4)}G_{B}^{(3)}G_{C}^{(3)}G_{D}^{(3)}G_{E}^{(3)}C_{F}^{(5)}
\notag \\
&&+\epsilon \mathcal{N}%
_{5}^{ijABCDE}G_{i}^{(5)}G_{j}^{(5)}G_{A}^{(3)}G_{B}^{(3)}G_{C}^{(3)}G_{D}^{(3)}C_{E}^{(5)}+\epsilon
\mathcal{O}%
_{5}^{ijkABCD}G_{i}^{(5)}G_{j}^{(4)}G_{k}^{(4)}G_{A}^{(3)}G_{B}^{(3)}G_{C}^{(3)}C_{D}^{(5)}
\notag \\
&&+\epsilon \mathcal{P}%
_{5}^{ijklABC}G_{i}^{(4)}G_{j}^{(4)}G_{k}^{(4)}G_{l}^{(4)}G_{A}^{(3)}G_{B}^{(3)}C_{C}^{(5)}+\epsilon
\mathcal{Q}%
_{5}^{iABCDEFG}G_{i}^{(4)}G_{A}^{(3)}G_{B}^{(3)}G_{C}^{(3)}G_{D}^{(3)}G_{E}^{(3)}G_{F}^{(3)}C_{G}^{(5)}.
\notag
\end{eqnarray}

\end{document}